\documentclass[aps,prl,amssymb,twocolumn,amsmath,groupedaddress]{revtex4-1}
\usepackage{graphicx}
\usepackage{dcolumn}
\usepackage{bm}

\begin{document}


\title{Two Classes of Singularities and Novel Topology in a Specially Designed Synthetic Photonic Crystals}


\author{Qiucui Li}
\author{Yu Zhang}
\author{Xunya Jiang}
\affiliation{Department of Illuminating Engineering and Light Sources, School of Information Science and Engineering,
Fudan University, Shanghai 200433, China.}
 \email[]{jinagxunya@fudan.edu.cn}


\date{\today}

\begin{abstract}
Zak phase and topological protected edge state are usually studied in one-dimensional(1D) photonic systems with spatial inversion symmetry(SIS). Interestingly in this work, we find specific classes of 1D structure without SIS can be mapped to system with SIS and also exhibit novel topology, which manifest as phase cut lines(PCLs) in our specially designed synthetic photonic crystals(SPCs). Zak phase defined in SIS is extended to depict the topology of PCLs after redefinition and topological protected edge state is also achieved in our 1D structure without SIS.
In our SPCs, the relationship between Chern numbers in two-dimension(2D) and the extended Zak phases of 1D PCLs is given, which are bound by the first type singularities. Higher Chern numbers and multi chiral edge states are achieved utilizing the concept of synthetic dimensions. The effective Hamiltonian is given, based on which we find that the band edges of each PCLs play a role analogous to the valley pseudospin and our SPC is actually a new type of valley photonic crystal which is usually studied in graphene-like honeycomb lattice. The chiral valley edge transport is also demonstrated. In higher dimension, the shift of the first type singularity in expanded parameter space will lead to the Weyl point topological transition which we proposed in our previous work. In this paper, we find a second type of singularities which manifests as a singular surface in our expanded parameter space. The shift of the singular surface will lead to the nodal line topological transition. Astonishingly, we find the states on the singular surface possess extremely high robust against certain randomness, based on which a topological wave filter is constructed.

\end{abstract}


\maketitle

\section{\label{sec:level1}INTRODUCTION}
In recent years, various novel topological phenomena addressed in the condensed matter physics\cite{PhysRevLett.45.494,RevModPhys.83.1057,RevModPhys.82.3045} have been achieved in photonic systems, where different types of topological insulators have all found their counterparts\cite{luling}. In one dimension, the photonic systems with SIS have been found to be direct analogs to Su-Schrieffer-Heeger model(SSH). The topology of the 1D systems with SIS is characterized by the Zak phase\cite{PhysRevLett.62.2747}. The connection between the Zak phases and the surface impedance in PCs is revealed\cite{PhysRevX.4.021017}. Topological protected edge states in 1D photonic system with SIS are constructed. The Zak phase of one isolated band is related to the singular point characterized by perfect transmission\cite{PhysRevX.4.021017,abc,PhysRevA.98.023838}. In two dimension, the PCs with Faraday-effect media serve as precise analogs to the quantum Hall effect systems, where chiral edge modes are extensively studied\cite{PhysRevLett.100.013904,PhysRevA.78.033834}. Spin Hall effect and valley Hall effect can also be achieved in conventional honeycomb lattice\cite{PhysRevB.96.201402,PhysRevLett.114.223901}. The topology index in two dimension is Chern number\cite{PhysRevLett.71.3697}. In three dimension, Weyl points have also been found in photonic system\cite{weyl,PhysRevLett.117.057401,PhysRevB.95.125136}.

On the other hand, the concept of synthetic system has attracted more and more interests, due to its ability of achieving novel physics in higher dimension and simplifying experimental designs\cite{PhysRevX.7.031032,Observation,PhysRevLett.113.050402,PhysRevLett.121.124501}. In two dimension, inspired by Aubry-Andr¨¦-Harper model(AAH)\cite{PhysRevLett.108.220401}, 2D topological band structure with nonzero Chern number has been realized in photonic AAH system and the topological radiative edge states exist in the gap characterized by nonzero reflection phase winding number\cite{PhysRevLett.112.107403,PhysRevA.91.043830,PhysRevB.93.195317}. In higher dimensions, the synthetic Weyl points and nodal lines are also constructed\cite{PhysRevX.7.031032,Observation,PhysRevLett.113.050402}. In our previous work\cite{2018arXiv181012550L}, utilizing a SPC we find that the high symmetric points where the system has SIS play a crucial role in determining the topology of different dimensions. The connection among the Zak phase of 1D systems with SIS, the Chern number in 2D and the topological phase transition with the Weyl points in higher-dimensional systems is revealed.

In this work, we find two classes of 1D structure without SIS can be mapped to certain 1D structure with SIS. Each class shares the same band structure, topology and similar wave function($E$ field) with their counterpart with SIS in our SPCs. They manifest as the PCLs with even and odd orders respectively in the reflection phase spectra. We propose that the Zak phase after redefinition can be extended to depict the topology of the PCLs where SIS is absent. The extended Zak phases of each PCLs are determined by the first type singularities which manifest as the vortex points in the reflection phase spectra.  We state that the first type singularities can only exist and shift along their corresponding PCLs. Based on the transfer matrix, we deduce that the opposite winding number of reflection phase vortex gives the topological charge of the corresponding singularities, which bind the extended Zak phases of 1D PCLs and the Chern numbers in 2D synthetic parameter space together. An explicit expression for the relationship between the extended Zak phases and the Chern numbers is given, based on which we find that the nonzero order PCLs will lead to higher Chern numbers. Based on the bulk edge correspondence the multi chiral edge states are demonstrated, which is a direct analog to quantum Hall effect. Based on the perturbation theory, the effective Hamiltonian is given for each gap. We find that Berry curvature sharply centralizes at the band edge points of each PCLs, which play a role of the high symmetric points of parameter space in our SPCs and we refer to as valleys. The states at the valleys are precise analogs of the valley pseudospin and our SPC is a new type of valley photonic crystal which is usually studied in graphene-like honeycomb lattice\cite{PhysRevB.96.201402,PhysRevLett.114.223901}. The chiral valley edge transport is guaranteed by the quantized projected valley Chern index. In higher dimension, we found in our previous work \cite{2018arXiv181012550L} the shift of the first type singularities in the expanded parameter space gives rise to the Weyl point topological transition. Here we find a second types of singularities which manifest as a singular line in the reflection phase spectra. When the singular line shifts from upper band to lower band, the corresponding gap will experience a nodal line topological transition and generate two pairs of new valleys with opposite Berry curvature. Chern numbers of the bands stay unchanged. We emphasize that the first type singularity is a singular point in the expanded parameter space, while the second type singularities form a singular surface in the expanded parameter space. In finite PCs, different from the trivial Bragg reflection induced perfect transmission, second type singularities exhibit extremely high robust against certain randomness, based on which we propose a topological wave filter.
\section{results}
\subsection{A.\label{A} Extended Zak                                                                                                                                             Phase In SIS Absent System and Topological Protected Edge State}
\begin{figure}[!htb]
\includegraphics[width=8.5cm]{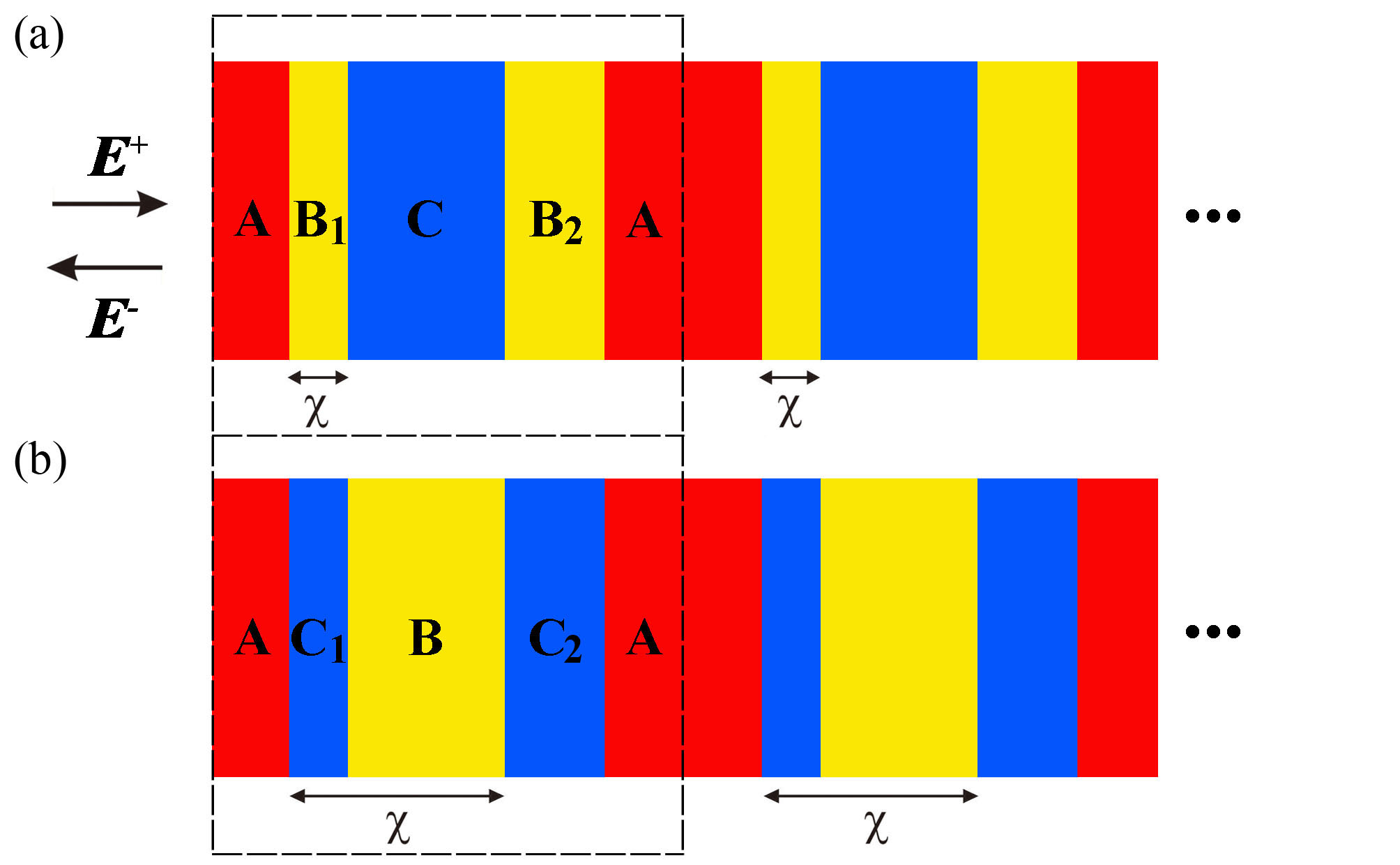}
\caption{\label{fig_1} The structure of our synthetic PC with $\chi$ within (a)$[0,d_b)$ (b) $[d_b,d_b+d_c)$. The plane wave normally incidents on it, whose forward and backward coefficients are denoted by $E^+$ and $E^-$. The unit cell is marked by black dashed box.}
\end{figure}
\begin{figure*}[!htb]
\includegraphics[width=17cm]{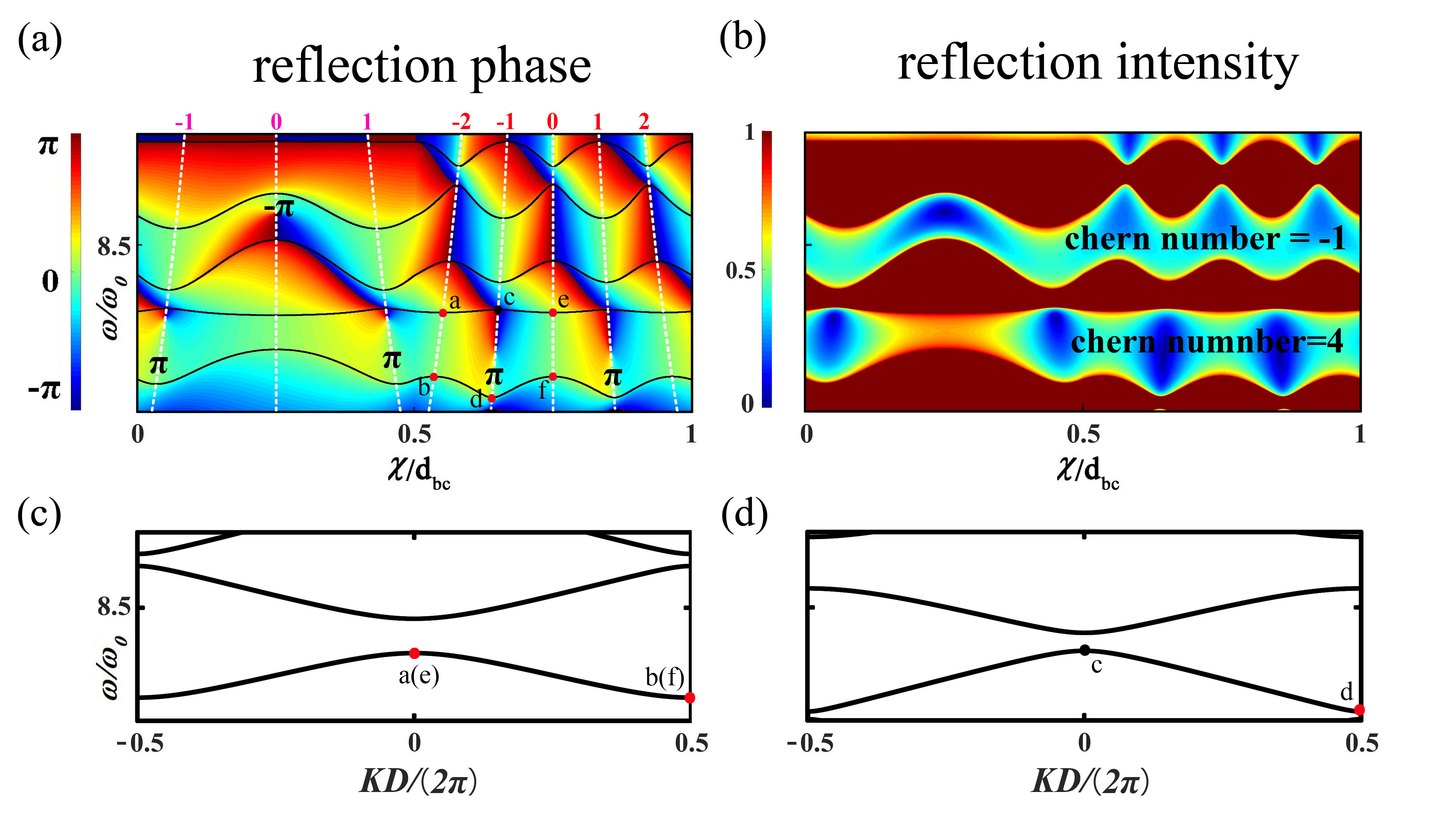}
\caption{\label{fig_2} The reflection phase $\phi$(a) and intensity $|r|$(b) as a function of frequency $\omega$ and synthetic dimension $\chi$. The parameters are given by $n_a=3.2n_b$, $n_c=2n_b$, $d_a=d_b=d_c=0.5d_{bc}$ and $\omega_0=c/(n_bd_b)$. The PCLs are marked by white dashed line, the order of which are labeled by red and pink numbers for $\chi$ within $[0,d_b)$ and $[d_b,d_b+d_c)$ respectively. Nonzero extended Zak phase of PCLs are labeled in(a) and Chern number of upper and lower bands are indicated in(b). The symmetric and antisymmetric band edge states are marked by red and black dots respectively, which are plotted in Fig.~\ref{fig_3}. (c) Band gap structure along $0$, $-2$ and $+2$ order PCLs with $\chi$ within $[d_b, d_b+d_c)$. (d) Band gap structure along $-1$ and $+1$ order PCLs with $\chi$ within $[d_b, d_b+d_c)$.}
\end{figure*}
We first consider a dielectric ACB-layered 1D PC as is shown in Fig.~\ref{fig_1}.with the
relative permittivity, relative permeability, refractive index, and width given by $\epsilon_i$, $\mu_i$, $n_i$, and $d_i$ respectively with $i=a,b$. The width of the unit cell is $D=d_a+d_b+d_c$.
To introduce the synthetic dimension, we continuously move the layer A across layer B and C from right to left. The distance between the right boundary of layer A and left boundary of layer C is denoted by $\chi$. When $\chi$ is between $0$ and $d_b$, the unit cell is $\rm{AB_1CB_2}$ and the width of left and right part of layer B are denoted by $d_{b1}$ and $d_{b2}$ respectively. When $\chi$ is between $d_b$ and $d_b+d_c$, the unit cell is $\rm{AC_1BC_2}$ and the width of left and right part of layer C are denoted by $d_{c1}$ and $d_{c2}$ respectively. When $\chi$ increases from $0$ to $d_b+d_c$, the system will transport around a loop. Bloch vector $K$ and $\chi$ construct a 2D closed parameter space in which Zak phase and related conclusions studied in SIS system\cite{PhysRevX.4.021017} will be extended to the present case without SIS. Electric-field $E_x$ in the layer A is given by:
\begin{equation}
E(z)=E^+\exp[ik_a(x+d_a/2)]+E^-\exp[-ik_a(x+d_a/2)]
\end{equation}
where $E^+$ and $E^-$ are the coefficients of forward and backward plane wave respectively and $k_a=wn_a/c$. The orgin is set to be the center of layer A. We define $E^-/(E^+exp(ik_ad_a))=|r|exp(i\phi)$ with the reflection intensity and  phase are denoted by $r$ and $\phi$ respectively. We notice that there is a special class of 1D system: the phase cut lines(PCLs) along which reflection phases equal to either $0$ or $\pi$ in the band as is shown in Fig.~\ref{fig_2} marked by white dashed lines. The existence of phase cut lines are given by:
\begin{equation}
\label{eq:2}
\begin{aligned}
&k_b(d_{b1}-d_{b2})=s\pi, \quad(0 \leq \chi \leq d_b) \qquad \\
&d_{b1}-d_{b2}=d_b-2\chi
\end{aligned}
\end{equation}
where $s \in \mathbb{Z}$ and represents the order of PCLs.
\begin{equation}
\label{eq:3}
\begin{aligned}
&k_c(d_{c1}-d_{c2})=p\pi, \quad (d_b \leq \chi \leq d_b+d_c) \\
&d_{c1}-d_{c2}=d_c+2d_b-2\chi
\end{aligned}
\end{equation}
where $p \in \mathbb{Z}$ and represents the order of PCLs.
We find that the zero order PCLs are the conventional 1D PC with spatial inversion symmetry(SIS) with the unit cell ABCB and ACBC when $\chi$ is within $[0,d_b)$ and $[d_b,d_b+d_c)$ respectively. Along the nonzero order PCLs, SIS is absent and the structure($\chi$) changes with the frequency, which we refer to as 1D PC with frequency dependent structure(PWFDS). For a certain point with frequency $\omega_0$ on the PCL, its $\chi_0$ can be calculated by Eq.~(\ref{eq:2}) and ~(\ref{eq:3}). The Bloch vector $K$ of the state$(\omega_0, \chi_0)$ can be calculated by the secular equation in Appendix A. In such a way the
$\omega$-$K$ dispersion relation(band gap structure) along each PCL can be constructed.

Next the geometry phase for each PCL can also be defined in a standard way. For the convenience of discussion, we focus on the PCLs characterized by Eq.~(\ref{eq:2}) of which $\chi$ is from $0$ to $d_b$. Along the $s$th order PCL, geometry phase $\theta^{+s}$ is expressed as: $\theta^{+s}=\int_{-\pi/D}^{\pi/D}[i\int{dx\epsilon(x)u^*(K,x)\partial_Ku(K,x)}]dK$ where $u(K,x)$ is the periodic part of the wave function\cite{PhysRevLett.62.2747}. As for the $0$th order pcl where $d_{b1}=d_{b2}$, the system has SIS of which geometry phase $\theta^{0}$ refered to as the Zak phase and topological protected edge states have been studied extensively in the previous work. As for the non-zero order PCL, since the SIS is absent, neither $\theta^{-s}$ nor $\theta^{+s}$ is well defined. However we can define $\theta^{s(-s)}_{Zak}=(\theta^{+s}+\theta^{-s})/2$ for the PCLs of order $+s$th and $-s$th which is either $\pi$ or $0$. Necessary mathematic prove is given in Appendix C.

\begin{figure}[!htb]
\includegraphics[width=8.5cm]{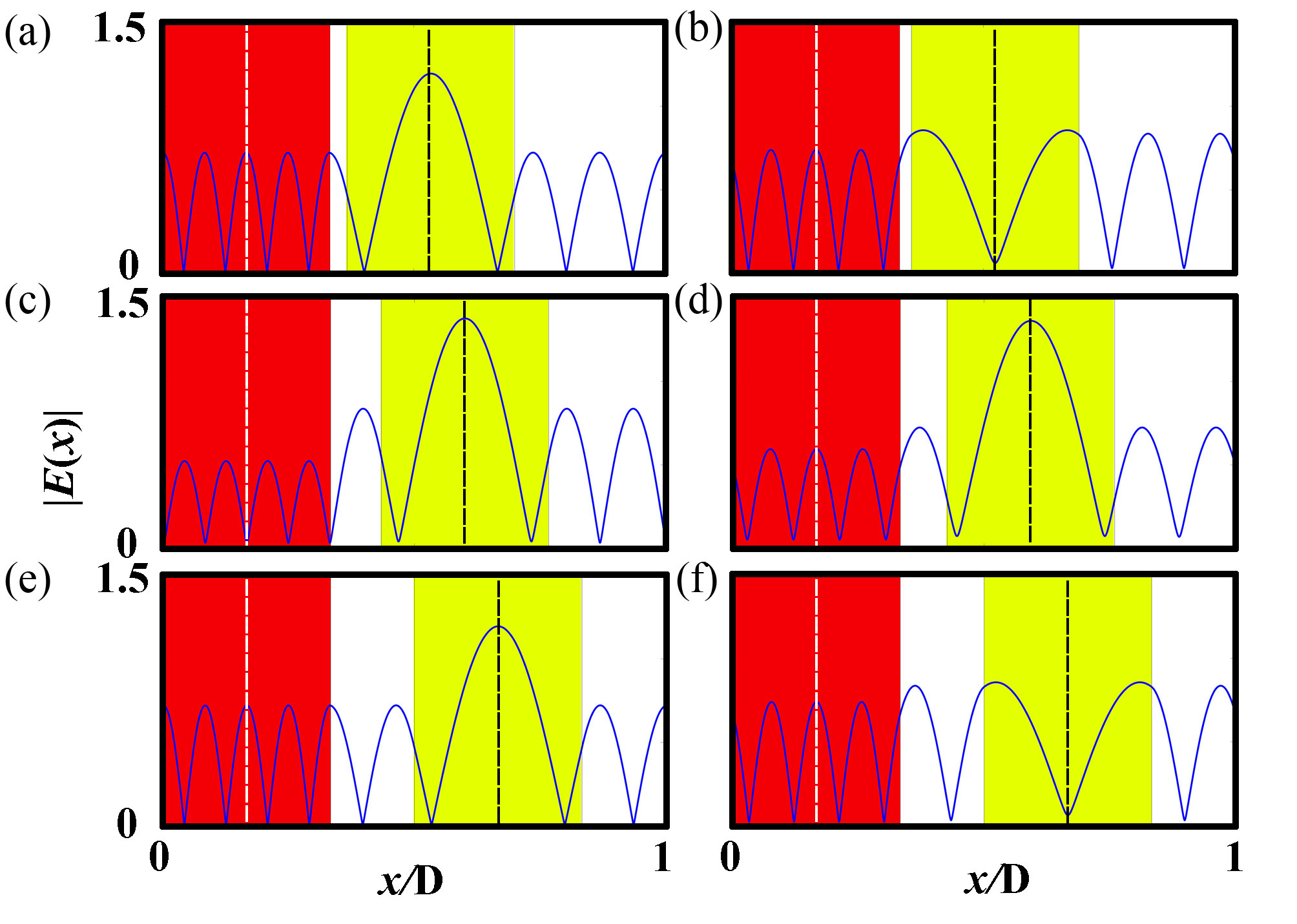}
\caption{\label{fig_3} Absolute value of $E$ field of six band edge states marked in Fig.~\ref{fig_2}. The red and yellow strips represent the layer A and B respectively. The center of the layer A and B are marked by white and black dashed lines respectively.}
\end{figure}

More interestingly, we find that the PCLs can be divided into two classes, even and odd order PCLs. All the even order PCLs have exactly the same $\omega$-$K$ dispersion relation(band gap structure), the same field inside layer A,C and the same extended Zak phases, which are clearly demonstrated in Fig.~\ref{fig_2} and ~\ref{fig_3}. So do the odd order PCLs.
Astonishingly we find that although the SIS is absent for nonzero order PCLs, these two classes PCLs can all be mapped into SIS system and the conclusions drawn in the system with SIS\cite{PhysRevX.4.021017} can be extended to depict PCLs without SIS after revision.

Firstly the extended Zak phase is entirely determined by the singularity at which the coefficients of forward and backward plane wave($E^+$ and $E^-$) turn to zero simultaneously and the Bloch states experience $\pi$ phase jump. The first type singularities manifest as a reflection phase vortex characterized by its zero reflection intensity as is shown in Fig.~\ref{fig_2}. They can only exist and transport along certain PCLs. For the PCLs of even order, The first type singularities are determined by:
\begin{eqnarray}
\label{eq:sepcl}
&&\tan{k_cd_c}=\frac{F_3\sin{k_bd_b}}{F_2\sin^2{k_bd_b/2-F_4\cos^2{k_bd_b/2}}},\nonumber\\
&&\sin{k_cd_c}\neq0
\end{eqnarray}
where
\begin{equation}
F_2=\frac{n_b^2}{n_cn_a}-\frac{n_cn_a}{n_b^2}, F_3=\frac{n_b}{n_a}-\frac{n_a}{n_b}, F_4=\frac{n_c}{n_b}-\frac{n_b}{n_c}
\end{equation}
For the PCLs of odd order, the existence of the first type singularity is given by:
\begin{eqnarray}
&&\tan{k_cd_c}=\frac{F_3\sin{k_bd_b}}{F_4\sin^2{k_bd_b/2-F_2\cos^2{k_bd_b/2}}},\nonumber\\
&&\sin{k_cd_c}\neq0
\end{eqnarray}
When there exist odd(even) numbers of singularities for a certain band, the extended Zak phase for the band will be $\pi$($0$). For the system with SIS, Zak phase can be determined by the parity of the band edge state at the high symmetric points of Brillouin zone. However, the SIS is absent for the non zero order PCLs and the band edge states don't have certain parity. We find that the symmetry of the band edge state inside
layer A determines the extended Zak phase for the non zero order PCLs. When the inside layer A of the two band edge states have the same(different) symmetry, the extended Zak phase will be $0$($\pi$) for the corresponding band. When the extended Zak phase equals to $0$($\pi$) relative to the center of layer A, it will be $\pi$($0$) relative to the center of layer B, which are clearly demonstrated in Fig.~\ref{fig_2} and ~\ref{fig_3}. Actually we state that the band edge points of each PCL not only determine the 1D topology along each PCL but also play a decisive role in the topology of higher dimensions in our synthetic system, which we refer to as the high symmetric points of Brillouin zone and denote as $P_0$ ,$P_1$ and $P_{-1}$ points etc according to the order of corresponding PCL.

\begin{figure}[!htb]
\includegraphics[width=8.5cm]{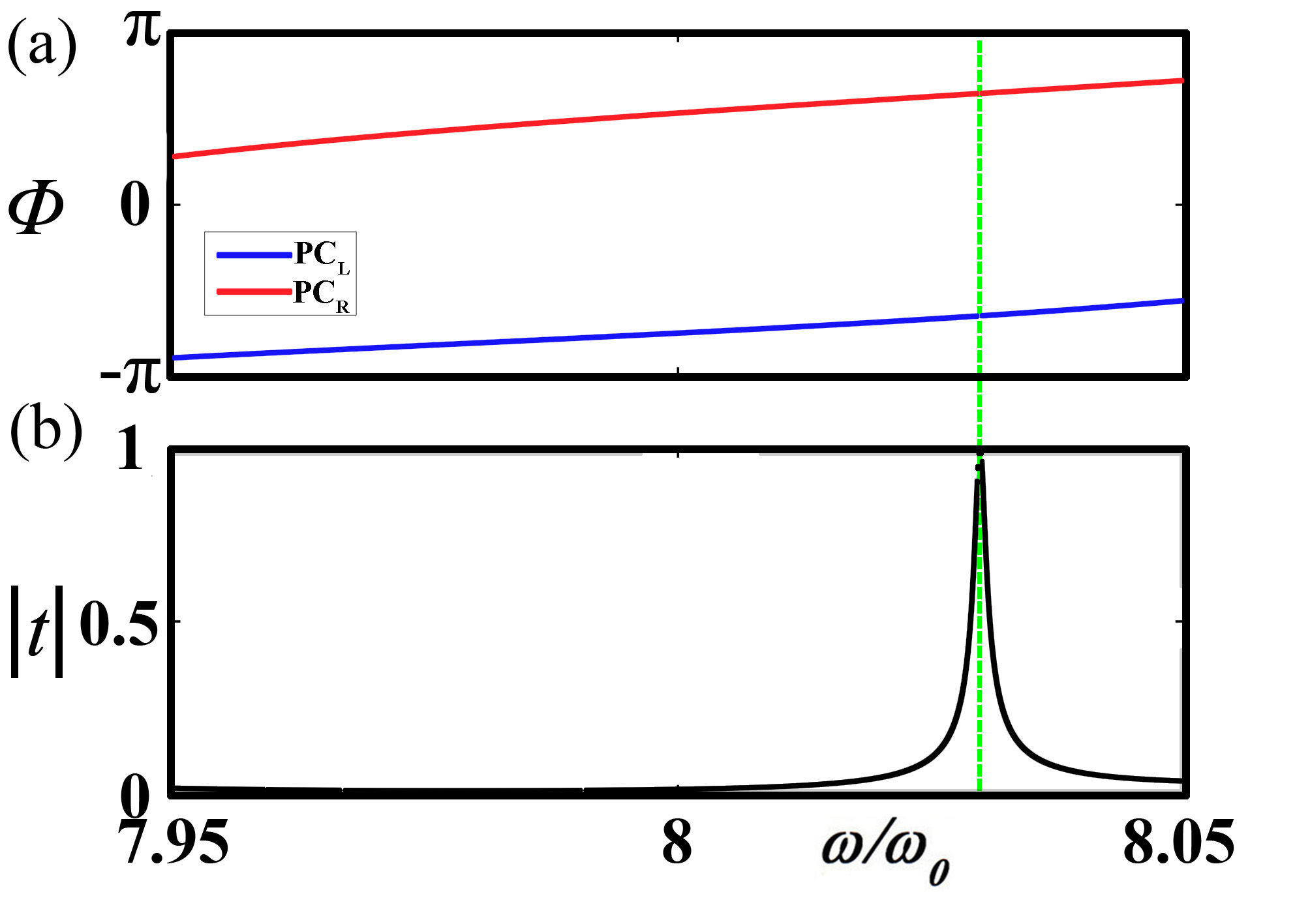}
\caption{\label{fig_4}The parameters are the same as Fig.~\ref{fig_2}. Reflection phase of left and right side PC along +1(-1) order PCL with $\chi$ within $[0,d_b)$ are plotted in blue and red line respectively in (a). Transmission intensity of attached PC with 10 cells on each side (b). Edge state is marked by green dashed line.}
\end{figure}

The topological protected edge state extensively studied in SIS system can also be constructed in our SIS absent system(non zero order PCLs).
Along sth order PCL, the sign of the reflection phase inside mth gap is determined by: sgn$[\phi_m]=(-1)^m(-1)^lexp(i\sum_{n=1}^m\theta^s_{Zak})$
where l is the number of band crossing points below the mth band gap, which is first proposed in the SIS system\cite{PhysRevX.4.021017}. The existence of the edge state at the interface between two semi-infinite 1D PCs attached together is given by:
\begin{equation}
\label{eq:11}
\phi_{L}+\phi_R=2j\pi,   j \in \mathbb{Z}
\end{equation}
where $\phi_{L}$ and $\phi_R$ are the reflection phase of the PCs at left and right side respectively. Inspired by the Zak phase inversion relative to the center of layer A and B, we construct the edge state by attaching two semi-infinite 1D PWFDS together. The boundary of the PCs at the left side is set to be the center of layer A and The boundary of the PCs at the right side is set to be the center of layer B. According to the relation between the extended Zak phase and the sign of the reflection phase, the mismatch of the Zak phase at left and right side PCs will lead to the opposite sign of $\phi_{L}$ and $\phi_R$ in the gap with odd index. So the edge state at the interface is guaranteed in the corresponding gap along non zero order PCLs. In Fig.~\ref{fig_4}, we make use of the $+1$(or$-1$) order PCL with $\chi$ within $[0,d_b)$ as is shown in Fig.~\ref{fig_2},
along which the reflection phase is shown in Fig.~\ref{fig_4}(a). The transmission intensity of the attached PC with 10 cells on each side is plotted in Fif.~\ref{fig_4}b, where the existence of the edge state is confirmed by the sharp peak marked by green dashed line. The frequency of the edge state is $\omega=8.03\omega_0$ where $\omega_0=c/(n_bd_b)$. According to Eq.~(\ref{eq:2}), $\chi$ of the attached PC is about $0.05$ and $4.45$ for the $-1$ and $+1$ order PCL respectively. In such a way the topological protected edge state in a special class of systems without SIS is constructed.

\subsection{B.\label{B} High Chern Numbers and Multi Edge States.}
In this section, we study the 2D system spanned by the Bloch vector $K$ and the synthetic dimension $\chi$. To characterise the topology of present 2D synthetic system, we introduce the Chern number in a standard way $\int_0^{d_b}d\chi\int_{-\pi/D}^{\pi/D}dK(\partial_KA_\chi-\partial_\chi{A_K})/(2{\pi}i)$, where $A_K=\int{dx\epsilon(x)u^*(K,x)\partial_Ku(K,x)}$. $u(K,x)$ is the periodic part of the wave function\cite{PhysRevLett.71.3697}.

Chern number of a certain band equals to the sum of the contour integral $\oint(d\chi{A_\chi}+dK{A_K})/(2{\pi}i)$ around all the first type singularities(reflection phase vortex) in this band, where $u_{K}$ is the periodic part of the E-field and $E_K(x) = u_K(x)\exp(iKx)$. We have deduced that the result of the contour integral around the phase vortex point equals to the opposite of the reflection phase winding number, which is deduced in Appendix D. The singularities give rise to not only non-zero extended Zak phase along different orders of PCLs, but also non-zero Chern number in our 2D synthetic system. In such a way, the singularities bind the 1D topology index(extended Zak phase) and the 2D topology index(Chern number) together in our 2D synthetic system.
In conclusion, the relationship between the Chern numebr($C_n$) of a certain band and the extend Zak phase can be expressed as:
\begin{equation}
\begin{aligned}
C_n=\sum_l s_{nl}\theta^l_{Zak}/\pi
\end{aligned}
\end{equation}
where $n$ and $l$ denote the $n$th band and order $l$th PCL respectively. When the winding number of the singularities are $-2\pi$ and $2\pi$, the $s_{nl}$ for the corresponding PCL are +1 and -1 respectively.

Analogue to the quantum Hall systems, the``bulk-edge correspondence" theory still works in our 2D synthetic systems that the number of the chiral edge states in a certain gap equals to the sum of Chern number of all bands below the corresponding gap. Such chiral edge states characterized by its unidirectional propagation can also find its counterpart in our 2D synthetic system. Due to the singularities of higher order PCLs, Higher Chern number of the band and multi edge states in the gap are achieved in our present system. As is shown in Fig.~\ref{fig_2}(a), the nonzero extended Zak phases along PCLs are indicated. We can see that the Chern numbers of upper and lower bands are $-1$ and $4$ respectively.

The existence of the edge state is also given by Eq.~(\ref{eq:11}). For the convenience of discussion, we assume our semi-infinite PC is attached to a perfect metal slab. Then the chiral edge states in the gap manifest as the $\pi$ reflection phase line which traverse the gap and connect the adjacent high symmetric points as is shown in Fig.~\ref{fig_2}(a). Actually, we can see that the winding number of the middle gap is four, no matter what reflecting substrate is, four chiral edge states can always be guaranteed according to Eq.~(\ref{eq:11}) in the gap. Our result can be confirmed by the measuring method of Chern number proposed in \cite{PhysRevA.91.043830}.
\begin{figure}[!htb]
\includegraphics[width=8.5cm]{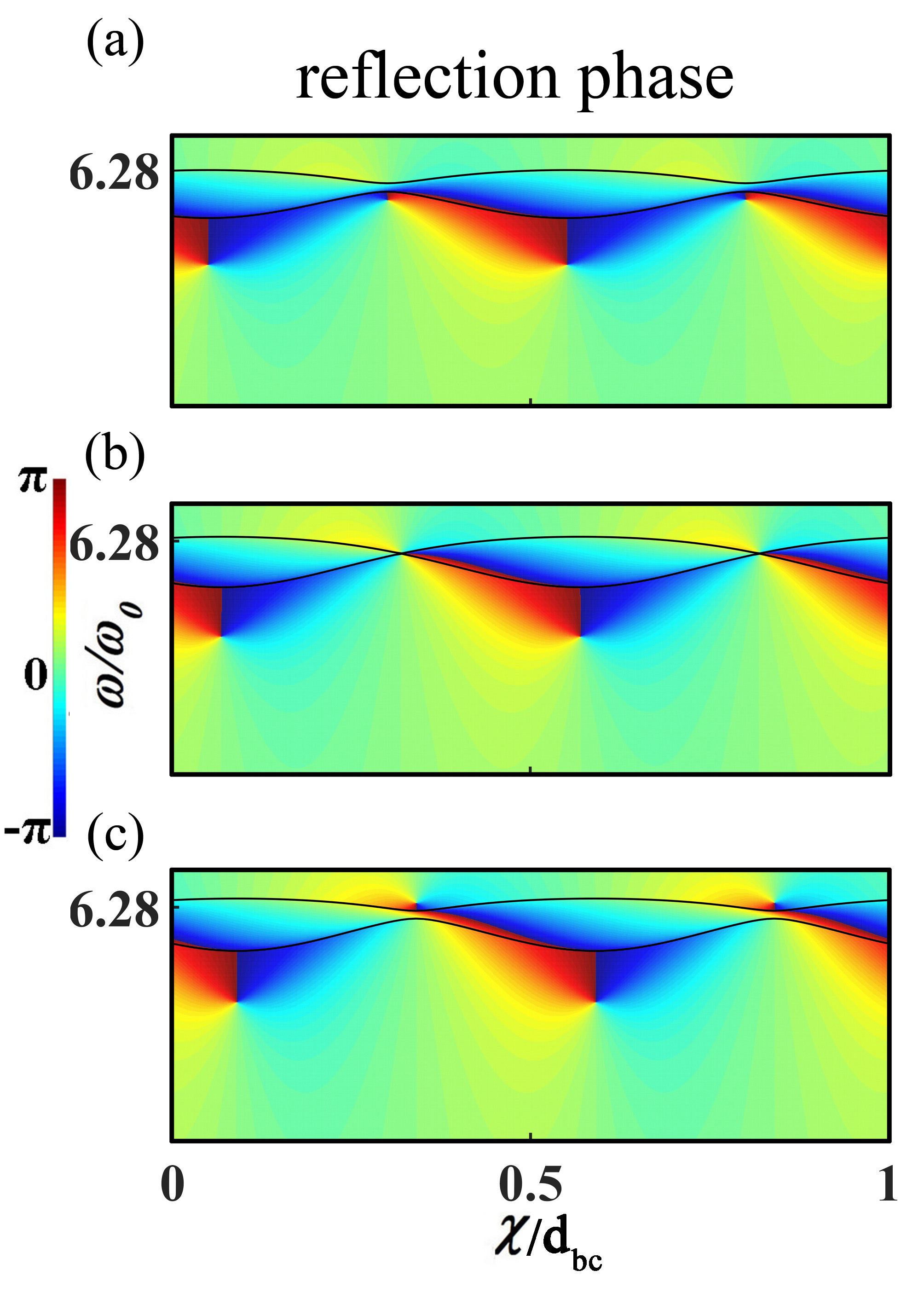}
\caption{\label{fig_5} Reflection phase spectra around the fourth gap is calculated by transfer matrix with $\epsilon_a=4\epsilon_b$, $\epsilon_c=1.01\epsilon_b$ and $d_a=0.4d_{bc}$ (a), $0.36d_{bc}$ (b), $0.32d_{bc}$ (c). Gap edges calculated by our effective Hamiltonian are plotted by black solid lines.}
\end{figure}
We notice that band inversion appears at the adjacent high symmetric points $P_0$ and $P_{1}$. The field inside layer A are symmetric and antisymmetric for the higher and lower frequency state at $P_0$ point, while at $P_1$ and $P_{-1}$ points the field inside layer A are antisymmetric and symmetric for the higher and lower frequency state. Such a symmetry switching process is the sufficient and necessary condition for constructing a topological nontrivial gap characterised by the existence of chiral edge state and bands with nonzero Chern numbers in our present system which we will discuss in detail.

\subsection{C.\label{c} Valley Hall Effect and Valley Chiral Edge State.}
More interestingly, when the switching process is absent and the band has zero Chern number, novel topology still exists in our synthetic system.
We will demonstrate that our synthetic system can serve as an valley photonic crystals as an analog of gapped valleytronic materials such as bilayer graphene. The states at the high symmetric points(band edge points of PCLs) serve as a direct analogue to the valley pseudospin which has found to be a new degree of freedom for electrons\cite{PhysRevB.87.155415}.
\begin{figure}[!htb]
\includegraphics[width=8.5cm]{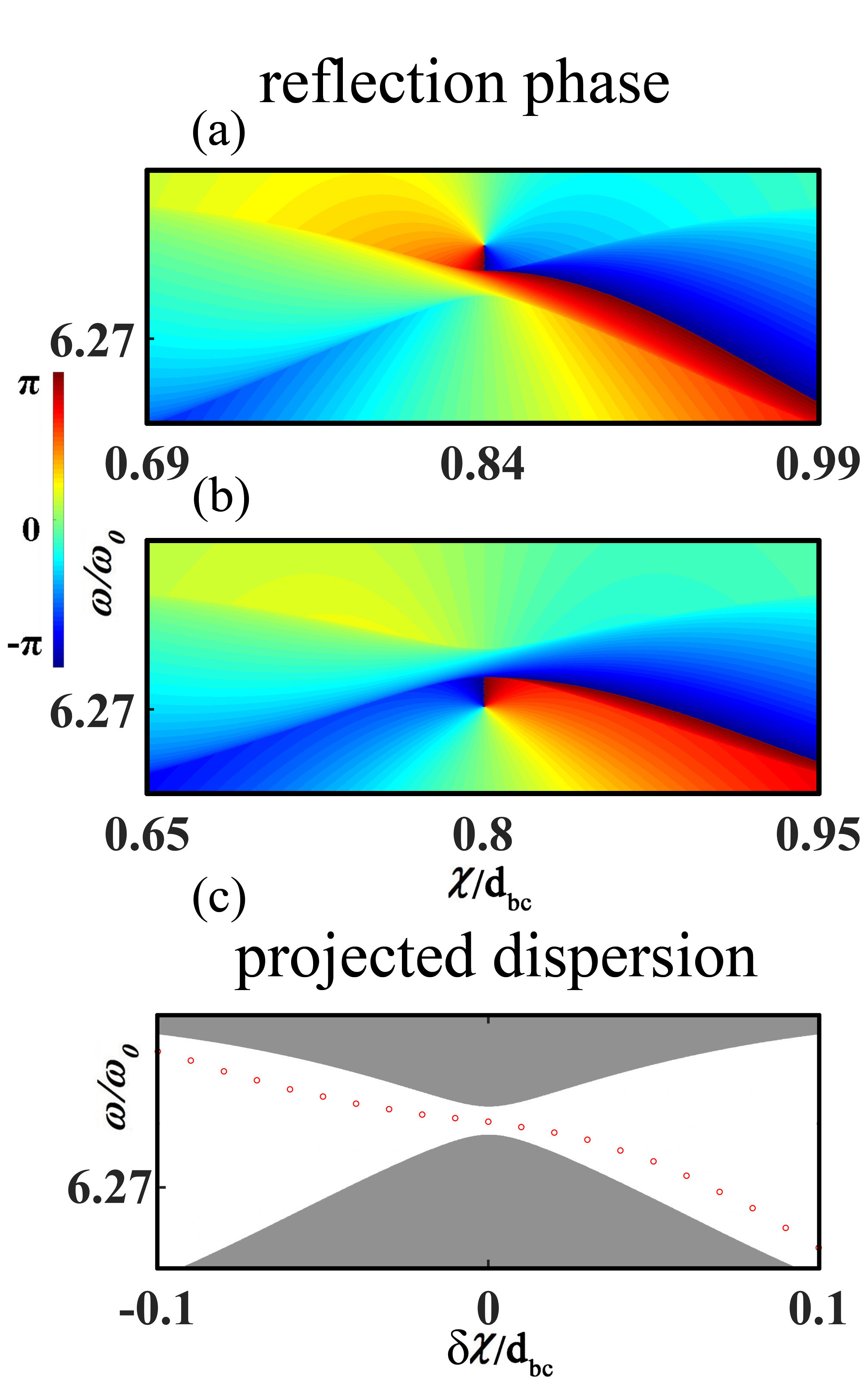}
\caption{\label{fig_6} Reflection phase spectra around valley $P_0$ with $n_a=2n_b$, $n_c=1.01n_b$, $d_a=0.5d_{bc}$ and $d_c=0.32d_{bc}$ $d_b=0.68d_{bc}$ for PC1 (a), $d_c=0.4d_{bc}$ $d_b=0.6d_{bc}$ for PC2 (b). Grey area in (c) represents the band of attached PC with 10 cells on each side. For certain $\delta\chi$, $\chi$ of the PCs on the left and right side are $\delta\chi+0.84d_{bc}$ and $\delta\chi+0.8d_{bc}$ respectively. The edge state is marked by red circle, which is confirmed by the transfer matrix.}
\end{figure}
We have derived the effective Hamiltonian based on the perturbation theory proposed in our previous work\cite{2018arXiv181012550L} to demonstrate the topology of different valleys. To derive the effective Hamiltonian, we treat the layer C as a perturbation layer and the deviation of $\epsilon_c$ from $\epsilon_b$ as a small number$\Delta\epsilon$. For the original unperturbed 1D AB-layered PC, we assume $n_ad_a = n_bd_{bc}$, so the $2n$th gap will closed at $K=0$ and $\omega=n\pi c/(d_a \sqrt{\epsilon_a})$, where $n \in \mathbb{N^+}$ and $d_{bc}=d_b+d_c$. We take the degenerated states with certain parity relative to the center of layer A as our basis.
The normalized symmetric state denoted by S-state:
\begin{subequations}
\label{eq:whole1}
\begin{equation}
-N_1k_a\cos{k_a(x-\frac{d_a}{2})},(0 \leq x \leq d_a) \label{subeq:1}
\end{equation}
\begin{equation}
N_1k_b\cos{k_a(x-d_a-\frac{d_{bc}}{2})},(d_a \leq x \leq d_{bc}+d_a) \label{subeq:2}
\end{equation}
\end{subequations}

\begin{figure}[!htb]
\includegraphics[width=8.5cm]{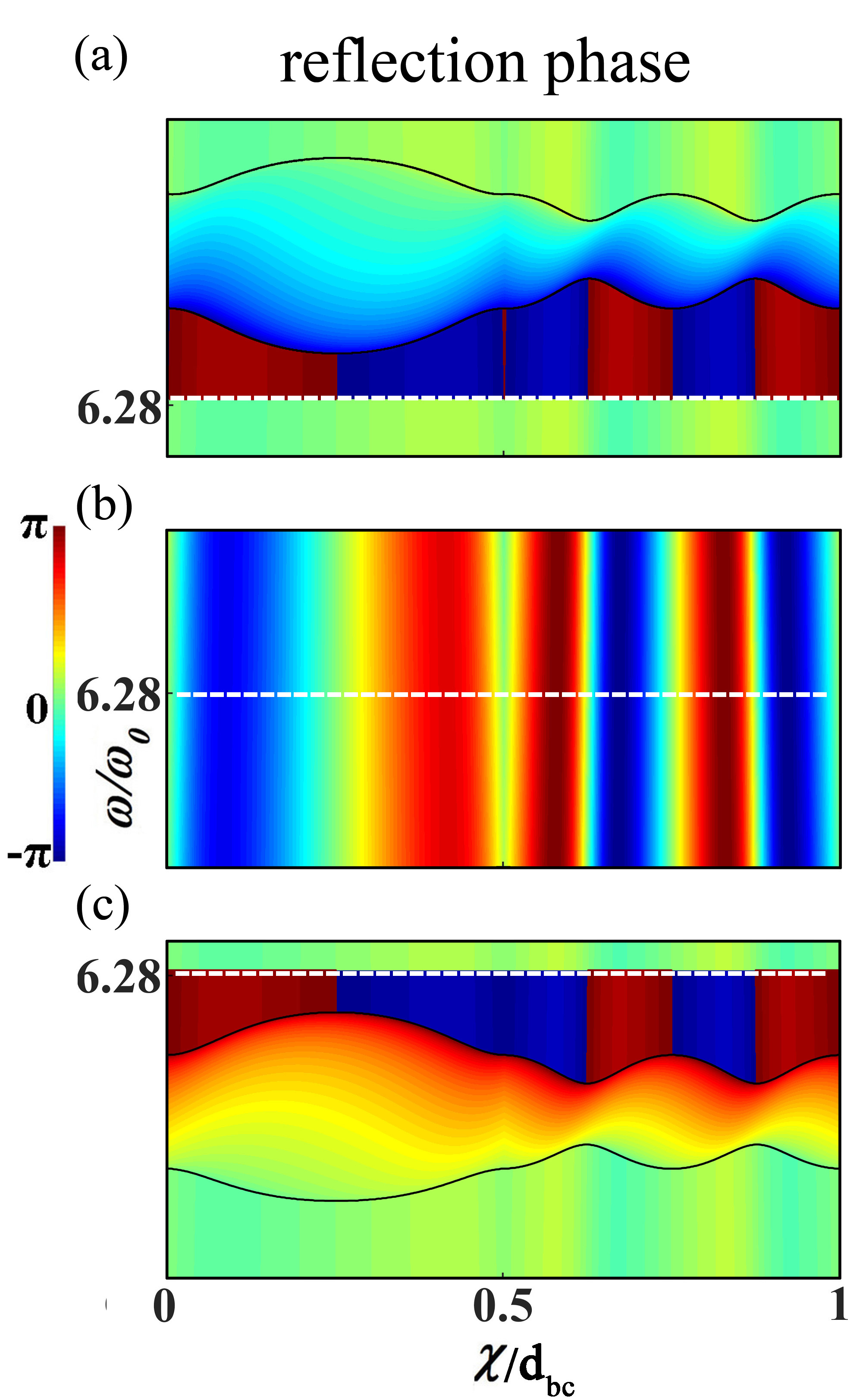}
\caption{\label{fig_7} Reflection phase spectra around $\omega=2\pi \omega_0$ with $d_a=d_b=d_c=0.5d_{bc}$ and $n_a=3.9n_b$ (a), $4n_b$ (b), $4.1n_b$ (c). The gap edges and singular lines are plotted by black solid and white dashed lines respectively. In (b), the gap is closed and turns into a degenerated line at $\omega=2\pi \omega_0$.}
\end{figure}

The normalized antisymmetric state denoted by A-state:
\begin{subequations}
\label{eq:whole2}
\begin{equation}
-N_2k_b\\sin{k_a(x-\frac{d_a}{2})},(0 \leq x \leq d_a) \label{subeq:3}
\end{equation}
\begin{equation}
N_2k_b\\sin{k_b(x-d_a-\frac{d_{bc}}{2})},(d_a \leq x \leq d_{bc}+d_a) \label{subeq:4}
\end{equation}
\end{subequations}
where $k_i = n_i\omega/c$, $n_i=\sqrt{\epsilon_i\mu_i}$, $N_1=1/\sqrt{k_a^2\frac{d_a}{2}+k_b^2\frac{d_{bc}}{2}}$ and $N_2=1/\sqrt{k_b^2\frac{d_a}{2}+k_b^2\frac{d_{bc}}{2}}$ are the normalized coefficients.

Followed the standard $\bm{K\cdot P}$ method, the Hamiltonian can be derived as:

\begin{equation}
H_{eff}=\alpha\sigma_0+\bm{\beta\cdot\sigma}
\end{equation}
where $\sigma_0$ is the identity matrix and $\bm{\sigma}$ is the Pauli matrix, while $\alpha$ and the vector $\bm{\beta_{K,\chi}}$ are defined as the coefficients of them:
\begin{equation}
\alpha=k_b^2+\frac{t(p(\chi)-2k_bd_c)N_1^2}{2}+\frac{t(-p(\chi)-2k_bd_c)N_2^2}{2}
\end{equation}
\begin{eqnarray}
\label{eq:9}
\bm{\beta}_{K,\chi}=[tq(\chi)N_1N_2,\frac{2k_bK}{\sqrt{c_1}},&&\frac{t(p(\chi)-2k_bd_c)N_1^2}{2}\nonumber \\
&&-\frac{t(-p(\chi)-2k_bd_c)N_2^2}{2}]
\end{eqnarray}
where
\[c_1=[(n_ad_a)^2+(n_bd_{bc})^2+(\frac{\sqrt{\epsilon_a}}{\sqrt{\epsilon_b}}+\frac{\sqrt{\epsilon_b}}{\sqrt{\epsilon_a}})n_ad_an_bd_{bc}]/D^2\],
t=$k_b^3\Delta\epsilon/4\epsilon_b^2$, and
\begin{subequations}
\begin{equation}
p(\chi)=\sin2k_b\chi-\sin2k_b(\chi-d_{bc}+d_c)\label{subeq:4}
\end{equation}
\begin{equation}
 q(\chi)=\cos2k_b(\chi-d_{bc}+d_c)-\cos2k_b\chi\label{subeq:4}
\end{equation}
\end{subequations}
Next we will derive the effective Hamiltonian to demonstrate the topology of different valleys. We take the fourth gap as an example with $k_ad_a = k_bd_{bc}=2\pi$. For the valley $P_0$ located at $K=0$ and $\chi=d_{bc}-d_c/2$, we expand the effective Hamiltonian with respect to $(\delta\chi,K)$ around the degenerated point where $\delta\chi=\chi-(d_{bc}-d_c/2)$:
\begin{equation}
\label{eq:dirac}
H_w=C\sigma_0+{\delta\chi}v_{\chi}\sigma_x+Kv_{K}\sigma_y+M\sigma_z
\end{equation}
where $C=k_b^2-t[(N_1^2-N_2^2)\sin{k_bd_c}+(N_1^2+N_2^2)k_bd_c]$, $v_{\chi}=-4N_1N_2k_bt\sin{k_bd_c}$, $v_{K}=2k_b/\sqrt{c_1}$ and $M=t[(N_2^2-N_1^2)k_bd_c-(N_1^2+N_2^2)\sin{k_bd_c}]$. When $d_c=0.3629d_{bc}$, $M=0$ and the gap is closed at the valley $P_0$, where a standard dirac point is formed according to Eq.~(\ref{eq:dirac}).
As depicted in Fig.~\ref{fig_5}, when $d_c<0.3629d_{bc}$, the reflection phase at the upper and lower edge of valley $P_0$ are $\pi$ and $0$ respectively.
So the state at the upper and lower edge of valley $P_0$ are A-state and S-state respectively. When $d_c>0.3629d_{bc}$, A-state and S-state are inverted at the adjacent valleys.
\begin{figure*}[!htb]
\includegraphics[width=17cm]{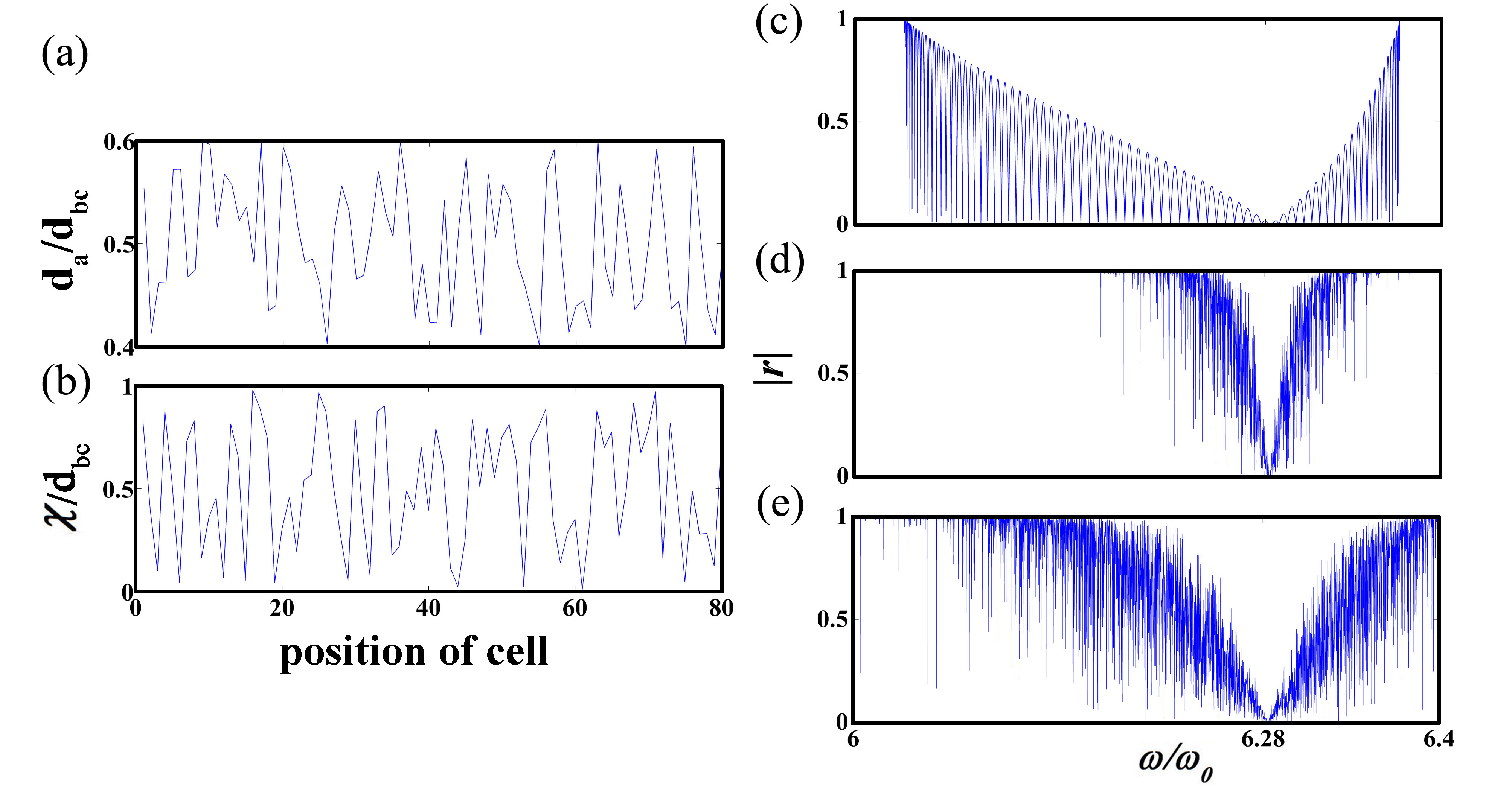}
\caption{\label{fig_8} The parameters are given by $n_a=3.7n_b, n_c=2n_b$, $d_b=d_c=0.5d_{bc}$ and N=80. (c) Reflection intensity of finite PC with unit cell ABCB and $d_a=0.5d_{bc}$. We first apply randomness to the width of layer A with $d_a=0.5(1+0.4w)d_{bc}$ and w is a random number between $[0,1)$, whose distribution with respect to position is plotted in (a). (d) The reflection intensity of the $d_a$ disordered PC. Next, we apply randomness to the $\chi$ of each cell, which is randomly distributed over $[0,1)$ with respect to position, whose distribution is shown in (b). (e) The reflection intensity of the $\chi$ disordered PC.}
\end{figure*}
For the valley $P_1$ located at $K=0$ and $\chi=d_{bc}-d_c/2+\pi/(2k_b)$, we expand the effective Hamiltonian with respect to $(\delta\chi,K)$, where $\delta\chi=\chi-(d_{bc}-d_c/2+\pi/(2k_b))$:
\begin{equation}
\label{eq:13}
H'_w=C'\sigma_0+{\delta\chi}v'_{\chi}\sigma_x+Kv'_{K}\sigma_y+M'\sigma_z
\end{equation}
where $C'=k_b^2-t[(N_2^2-N_1^2)\sin{k_bd_c}+(N_1^2+N_2^2)k_bd_c]$, $v'_{\chi}=4N_1N_2k_bt\sin{k_bd_c}$, $v'_{K}=2k_b/\sqrt{c_1}$ and $M'=t[(N_2^2-N_1^2)k_bd_c+(N_1^2+N_2^2)\sin{k_bd_c}]$.
The Hamiltonian Eq.~(\ref{eq:dirac}) and (\ref{eq:13}) indicate a nontrivial valley-dependent Berry curvature which has a distribution sharply centered at the valleys. More interestingly, we notice that the sign of $v_{\chi}$(the coefficient of $\sigma_x$) is opposite at the adjacent valleys. So if the sign of $M$(the coefficient of $\sigma_z$) is unchanged at the adjacent valleys, the Berry curvature will be opposite at the adjacent valleys and the corresponding gap will contribute $0$ to the Chern number of the upper and lower bands. In such a way we propose that the sufficient and necessary condition to introduce non zero Chern number is the existence of  symmetry switching process at the adjacent valleys.
\begin{figure}[!htb]
\includegraphics[width=8.5cm]{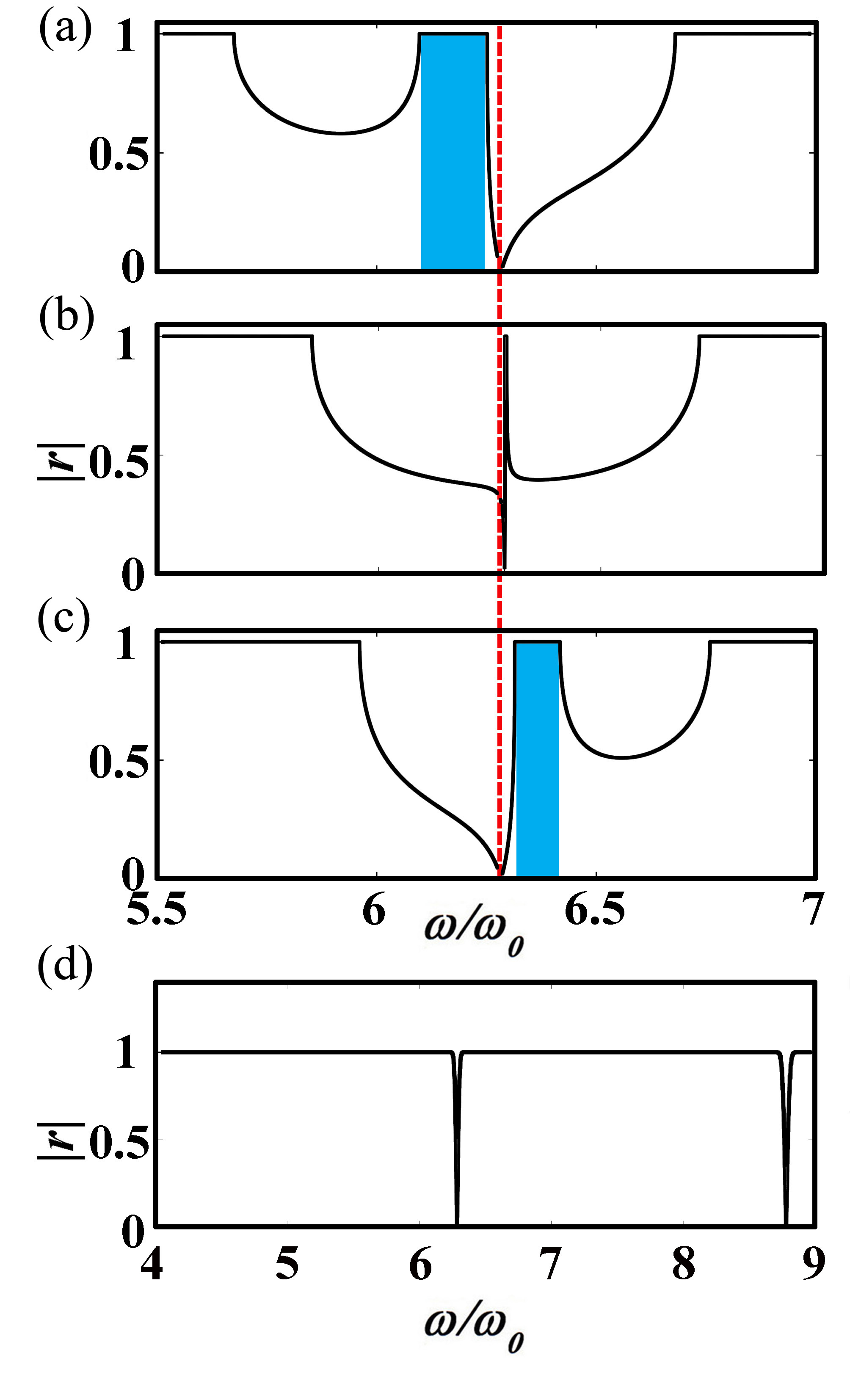}
\caption{\label{fig_9} The parameters are given by $n_a=3.7n_b, n_c=2n_b$ and $d_b=d_c=0.5d_{bc}$. Reflection intensity of semi-infinite PC composed of with unit cell ABCB with $d_a=0.42d_{bc}$ (a), $d_a=0.54d_{bc}$ (a) and $d_a=0.66d_{bc}$ (c). Reflection intensity of topological wave filter composed of 80 cells with $d_a$ gradually increasing from $0.42d_{bc}$ to $0.66d_{bc}$.}
\end{figure}
In summary, when $d_c<0.3629d_{bc}$, the symmetry switching process exists at the adjacent valleys $P_0$ and $P_1$, the Chern number of the lower band is $2$ and the fourth gap is topological nontrivial with two chiral edge states as is shown in Fig.~\ref{fig_5}(c), which can be seen as a direct analogue to quantum Hall effect. In this case, the sign of the Berry curvature at valley $P_0$ and $P_1$ are the same.

when $d_c>0.3629d_{bc}$, although the symmetry switching process is absent and the total Chern number of the lower band is $0$, topology of the fourth gap  can be characterised by valley Chern index. In this case, the sign of the Berry curvature at valley $P_0$ and $P_1$ are opposite, which give rise to the valley Chern index $C_{P0(P1)}$ that takes the value of $\mp1/2$. The adjacent valleys with opposite Berry curvature are direct analogs to the pseudospin-up and pseudospin-down in the valley Hall effect system\cite{PhysRevB.96.201402}. The chirality of the certain valley can be characterised by measuring the reflection phase's winding number of the singularity at the corresponding PCL. For example, the winding number of the singularity is $1$ at the $0$th order PCl, which give rise to the valley Chern index $C_{P0}=-1/2$ for the valley $P_0$. the winding number of the singularity is $-1$ at the $1$th order PCl, which give rise to the valley Chern index $C_{P1}=1/2$ for the valley $P_1$.
When passing through a topological transition point, the singularity transport from the lower band to the upper band along a phase cute line and the valley Chern index of the certain valley is inverted at the same time.

The valley edge transport can also be achieved in our present system by attaching two 1D PCs denoted by PC1 and PC2. We set $n_a=2n_b$ and $n_c=1.01n_b$. As discussed above, when $d_a=0.5d_{bc}$, $d_c=0.3629d_{bc}$ and $d_b=0.6371d_{bc}$, the gap will be closed and turn into a dirac point. In the Fig.~\ref{fig_6}, we set $d_c=0.32d_{bc}$ and $d_c=0.4d_{bc}$ for PC1 and PC2 respectively. At the same time, we also tune $d_b$ in such a way as to keep $n_ad_a+n_bd_b+n_cd_c$ unchanged so that the frequency of the mid-gap is maintained. The valley Chern index of PC1 and PC2 is opposite because of passing through a topological transition point. We construct a interface between PC1 and PC2.  At the coordinate origin of the projected dispersion Fig.~\ref{fig_6}(c), the $\chi$ of PC1 and PC2 are set to be $\chi_1=0.84d_{bc}$ and $\chi_2=0.8d_{bc}$(the position of their valley) respectively so as to project the valley $P_0$ of PC1 onto the corresponding valley of PC2 and $\delta\chi=\chi_1-0.84d_{bc}=\chi_2-0.8d_{bc}$. The difference in the valley projected topological charges is quantized across the interface $\left |\Delta C_{P0}=C_{P0}^1-C_{P0}^2\right |=1$, where $C_{P0}^1$ and $C_{P0}^2$ are the $P_0$ valley Chern index of the PC1 and PC2 respectively, which guarantees the existence of a chiral edge state for this valley, as is shown in Fig.~\ref{fig_6}(c), which is a direct analog to the valley quantum Hall effect.
\subsection{D.\label{d} Two Different Types Singularities Induced Weyl Points and Nodal Lines in Higher Dimensions}

In this section, we introduce $d_c$(the width of layer C) as a new variable as we did in our previous work\cite{2018arXiv181012550L}.
we find that the gap will be closed when $dc$ satisfies $(N_1^2-N_2^2)k_bd_c+(N_1^2+N_2^2)\sin{k_bd_c}=0$. We obtained $d_c=d_0=0.3629d_{bc}$.
\begin{equation}
\label{eq:weyl}
H_w=(C+D\delta{d})\sigma_0+{\delta\chi}v_{\chi}\sigma_x+Kv_{K}\sigma_y+\delta{d}{v_d}\sigma_z
\end{equation}
where $C=k_b^2-t[(N_1^2-N_2^2)\sin{k_bd_0}+(N_1^2+N_2^2)k_bd_0]$, $D=-k_bt[(N_1^2-N_2^2)\cos{k_bd_0}+(N_1^2+N_2^2)]$, $v_{\chi}=-4N_1N_2k_bt\sin{k_bd_0}$, $v_{K}=2k_b/\sqrt{c_1}$ and $v_d=k_bt[N_2^2-N_1^2-\cos{k_bd_0}(N_1^2+N_2^2)]$. When $\delta{d}$ is tuned from positive to negative, one singularity with reflection phase winding number $2\pi$ moves from lower band to upper band along the $0$th order PCL and at the topological transition point the band degenerate point is actually a Weyl point characterised by the standard Weyl Hamiltonian in the expanded parameter space$\{\chi,K,d_c\}$, which serves as the the instantaneous jumping channel of the singularity between two bands. The charge of this Weyl point is one according to standard definition\cite{weyl}. Such a Weyl point indicates the existence of other Weyl points with the same topological charge in all valleys with even order since band structure and topology are identical along the PCLs of even order.

There is an another type of singularity. The second type singularities manifest as a singular line in the reflection phase spectra at the frequency $\omega'$ determined by $\sin n_bd_b \omega'/c=0$ and $\sin n_cd_c \omega'/c=0$. We notice that the frequency of the second type singularity(singular line) is entirely determined by the optical path ratio of layer B and layer C($n_bd_b/n_cd_c$) and independent of $\chi$ and $n_ad_a$. For all the states at the singular line, reflection phase experiences $\pi$ phase jump and reflection intensity reaches $1$. As is shown in Fig.~\ref{fig_7}, the frequency of the degenerate line is $\omega'=2\pi c/n_bd_b$. We continuously tune $n_ad_a$ from $3.9n_bd_b$ to $4.1n_bd_b$, the singular line moves from lower band to upper band. At the critical point $n_ad_a=4n_bd_b$, the singular line reaches the frequency of the mid-gap and the gap turns into a degenerate line marked by the white dashed line. The movement of the singular line in the expanded parameter space serves as the direct origin of the nodal line band structure in higher dimension for our synthetic system. The most general condition of nodal lines in our model is that:
We assume that the optical path ratio of layers satisfies
\begin{equation}
n_a d_a:n_b d_b:n_c d_c=l:m:n
\end{equation}
where $l, m, n \in \mathbb{N^+}$. Then the $l+m+n$th gap will turn into a nodal line at the frequency $w_{l+m+n}=(l+m+n)\pi{c}/(n_ad_a+n_bd_b+n_cd_c)$. See our Appendix E.
We state that the second type singularities do not contribute to the Chern number of the band. When a singular line passing through a certain gap, the winding number of the gap and topology of the adjacent bands stay unchanged. According to Eq.~(\ref{eq:2}) and ~(\ref{eq:3}), we notice that the starting frequency of $\pm s$th order PCL is given by $w_s=s\pi c/(n_bd_b)$ with $\chi$ within $[0,d_b)$ and the starting frequency of $\pm p$th order PCL is given by $w_p=p\pi c/(n_bd_b)$ with $\chi$ within $[d_b,d_b+d_c)$. So we can see that four new PCLs are generated across the singular line. More interestingly, when the band structure passing through a nodal line topological transition with the singular line transferring from upper band to the lower band, the corresponding gap will generate four new valleys with one pair within $[0,d_b)$ and the other pair within $[d_b,d_b+d_c)$. The Berry curvature of the two pair valleys are opposite, so the topology of the gap stay unchanged.

\subsection{E.\label{e} Extremely High Robust Singular Surface and Topological Wave Filter}

We find that the second type singularities exhibit extremely high robust against randomness. We first assume that our 1D synthetic PC consists of $80$ cells. As is shown in Fig.~\ref{fig_8}., the reflection intensity exhibits a sequence of zero points which arises from the Bragg reflections at the Bloch vector $K$ satisfying $NKD=m\pi$ where $m=0,1,...,N-1$ and $N$ is the number of cells. Aside from these trivial zero point additional zero point refers to our second type singularity appearing at the frequency $\omega'=2\pi c/(n_bd_b)$. Next we assume the $\chi$ of each cell is randomly distributed between $[0,1)$ as is shown in Fig.~\ref{fig_8}(b), we find that the frequency of our second type singularity stays unchanged and exhibits extremely high robust against randomness of $\chi$ while the trivial zero points caused by Bragg reflection are all annihilated by the randomness. We also investigate the influence of $d_a$ randomness. We assume $d_a$ of each cell has a disorder strength of $0.4$, which is expressed as $d_a=0.5(1+0.4w)d_{bc}$ where w is a random number between $-0.5$ and $0.5$. The robust against $d_a$ randomness is also confirmed for our second type singularity with its frequency unchanged as is shown in Fig.~\ref{fig_8}. Actually in the parameter space $\{ n_ad_a, K, \chi \}$, our second type singularity actually is a singular surface with its frequency independent of $\chi$ and $n_ad_a$. The high robust are shared by the point of the singular surface.

Next we will construct a topological wave filter by making use of our high robust second type singularity. To demonstrate our idea, we continuously tune $d_a$ from $0.42d_{bc}$ to $0.66d_{bc}$. The reflection intensity of semi-infinite 1D PC is shown in Fig.~\ref{fig_9}. We notice that the topological transition occurs at the gap marked by cyan strip which is closed at $d_a=0.54d_{bc}$ and reopened. The mid gap frequency passes through the frequency of the singularity marked by the red dashed line which stays unchanged in the whole process. We assume that a finite ABCB layered 1D PC consists of $80$ cells with $d_a$ gradually varying from $0.42d_{bc}$ to $0.66d_{bc}$. The moving gap erases all the frequency except for the robust second type singularity. The reflection intensity of our topological wave filter as is shown in Fig.~\ref{fig_9}(c) exhibits a sharp peak at the frequency of singularity.

In summary, we find new classes of 1D structure(PCLs) without SIS can be mapped to system with SIS, whose topology can be depicted by the redefined Zak phase. The topological protected edge states are also constructed in such SIS-absent 1D structure. The extended Zak phase is determined by the first type singularity. The relationship between the topological charge of the first type singularity in the 2D expanded parameter space and the winding number of the reflection phase vortex is revealed, based on which the extended Zak phase and the Chern number are related. Based on the effective Hamiltonian, we reveal that our SPC serves as a new type of valley PC which is extensively studied in honeycomb lattice. Band edge points of PCLs play a role analogous to the valley pseudospin. Multi chiral edge states and chiral edge transport are demonstrated. In expanded higher dimension, the first type singularity manifests as a singular point, which serves as the origin of Weyl point band structure. A second type singularity is found, which manifests as a singular surface in the expanded parameter space. Passing through a nodal line topological transition, new valleys will be generated(annihilated) but the Chern numbers of each band stay unchanged. The extremely high robust of the second type singularity is shown, based on which we propose a topological wave filter.

\section*{ACKNOWLEDGMENTS}
This work was supported by a grant from the National Nature Science Foundation of China(11334015), a grant from the National Key Research Program of China (2016YFA0301103,2018YFA0306201) and National High Technology Research and Development Program of China(863Progtam)(17-H863-04-ZT-001-035-01).

\section*{Appendix A: \label{A}Transfer matrix and dispersion relation}
The eigen equation for our synthetic system can be written as:
\begin{eqnarray}
\label{eq:S1}
[{\bf T}-\exp (iKD)]
\left(
\begin{array}{c}
E^+\\
E^-
\end{array}\right)\;
=0
\end{eqnarray}
{\bf T} is the transfer matrix for the unit cell. $K$ and $D$ are the Bloch vector and width of the unit cell respectively. $E^+$ and $E^-$ are the coefficients of the forward and backward-propagating waves inside layer A respectively. We have:
\begin{equation}
\label{eq:efa}
\begin{aligned}
&E^+=T_{12} \\
&E^-=\exp(iKD)-T_{11}
\end{aligned}
\end{equation}
The electric field inside layer A can be written as:
\begin{equation}
E(z)=E^+\exp[ik_a(z+d_a/2)]+E^-\exp[-ik_a(z+d_a/2)]
\end{equation}
with the origin set to be the center of layer A. The reflection phase at the center of layer A is:
\begin{equation}
\label{eq:rf}
r=\frac{\exp(iKD)-T_{11}}{T_{12}\exp(ik_ad_a)}
\end{equation}
Without loss of generality, we assume $\chi$ is from $0$ to $d_b$ and the unit cell is $\rm{AB_1CB_2}$. The transfer matrix is given by:
\begin{widetext}
\begin{eqnarray}
\label{eq:rt11}
Re[T_{11}]=&&\cos{k_ad_a}\cos{k_cd_c}\cos{k_bd_b}-\frac{1}{2}M_1\cos{k_bd_{b1}}\cos{k_bd_{b2}}\sin{k_ad_a}\sin{k_cd_c} \nonumber\\
&&+\frac{1}{2}M_2\sin{k_bd_{b1}}\sin{k_bd_{b2}}\sin{k_ad_a}\sin{k_cd_c}-\frac{1}{2}M_3\sin{k_bd_b}\sin{k_ad_a}\cos{k_cd_c}\nonumber\\
&&+\frac{1}{2}M_4\sin{k_bd_b}\cos{k_ad_a}\sin{k_cd_c}
\end{eqnarray}
\begin{eqnarray}
\label{eq:it11}
Im[T_{11}]=&&\sin{k_ad_a}\cos{k_cd_c}\cos{k_bd_b}+\frac{1}{2}M_1\cos{k_bd_{b1}}\cos{k_bd_{b2}}\cos{k_ad_a}\sin{k_cd_c} \nonumber\\
&&-\frac{1}{2}M_2\sin{k_bd_{b1}}\sin{k_bd_{b2}}\cos{k_ad_a}\sin{k_cd_c}+\frac{1}{2}M_3\sin{k_bd_b}\cos{k_ad_a}\cos{k_cd_c}\nonumber\\
&&-\frac{1}{2}M_4\sin{k_bd_b}\sin{k_ad_a}\sin{k_cd_c}
\end{eqnarray}
\begin{eqnarray}
\label{eq:t12}
T_{12}=&&\frac{e^{-ik_ad_a}}{2}[iF_1\sin{k_cd_c}\cos{k_bd_{b1}}\cos{k_bd_{b2}}-iF_2\sin{k_cd_c}\sin{k_bd_{b1}}\sin{k_bd_{b2}}\nonumber\\
&&+iF_3\cos{k_cd_c}\sin{k_bd_b}-F_4\sin{k_cd_c}\sin{k_b(d_{b2}-d_{b1})}]
\end{eqnarray}
where
\begin{equation}
M_1=\frac{n_a}{n_c}+\frac{n_c}{n_a}, M_2=\frac{n_b^2}{n_cn_a}+\frac{n_cn_a}{n_b^2}, M_3=\frac{n_a}{n_b}+\frac{n_b}{n_a},
M_4=\frac{n_b}{n_c}+\frac{n_c}{n_b}
\end{equation}
and
\begin{equation}
F_1=\frac{n_c}{n_a}-\frac{n_a}{n_c}, F_2=\frac{n_b^2}{n_cn_a}-\frac{n_cn_a}{n_b^2}, F_3=\frac{n_b}{n_a}-\frac{n_a}{n_b}, F_4=\frac{n_c}{n_b}-\frac{n_b}{n_c}
\end{equation}
\end{widetext}
The dispersion relation is give by:
\begin{equation}
\label{eq:dr}
Re[T_{11}]=\cos{KD}
\end{equation}
\section*{Appendix B: \label{B}Phase cut lines}
When $k_b(d_{b1}-d_{b2})=n\pi, n \in \mathbb{Z}$, $T_{12}\exp(ik_ad_a)$ is a pure imaginary number. Since $\exp(iKD)-T_{11}$ is also a pure imaginary number, $k_b(d_{b1}-d_{b2})=n\pi, n \in \mathbb{Z}$ gives the existence of phase cut lines of order n, along which the reflection phase $\phi$ is either $0$ or $\pi$. The starting frequency $\omega_n$ of PCLs with order n is given by $\omega_n=n\pi c/n_bd_b$.

The PCLs of even order is given by:
\begin{equation}
\label{eq:pcl}
k_b(d_{b1}-d_{b2})=2l\pi, l \in \mathbb{Z}
\end{equation}
Substituting Eq.~(\ref{eq:pcl}) into Eq.~(\ref{eq:dr}) and ~(\ref{eq:efa}), we can find that all the even order PCLs have exactly the same dispersion relationship(band-gap structure) and same field inside layer A. By switching $d_a$ and $d_c$, we can see that the field inside layer C is also the same along all the even order PCLs. At the band edge point of each PCLs, the field inside layer A and C are either symmetric or antisymmetric, or we can state equivalently that the amplitude of the wave function at the center of layer A and C is either zero or maximum.

\section*{Appendix C: \label{C}First type singularities and extended Zak phase}
The condition of singular point at which $E^+=0$ and $E^-=0$ simultaneously is given by $E^+=T_{12}\exp(ik_ad_a)=0$ and they all exits in the band branch with positive group velocity $\partial\omega/\partial{K}\geq0$ under the gauge defined in this Appendix. At the band branch with positive group velocity since $|E^+|\geq|E^-|$, $E^+=T_{12}\exp(ik_ad_a)=0$ will lead to $E^-=0$ at the same time. As for a certain singular point located at $\omega$ and $K$, we emphasize that for the state with same frequency and opposite Bloch vector\{$\omega$,$-K$\}, $E^+=0$ but $E^-\neq0$ which does not act as a singular point.

For the PCLs of even order, $T_{12}=0$ gives:
\begin{eqnarray}
\label{eq:sepcl}
&&\tan{k_cd_c}=\frac{F_3\sin{k_bd_b}}{F_2\sin^2{k_bd_b/2-F_4\cos^2{k_bd_b/2}}},\nonumber\\
&&\sin{k_cd_c}\neq0
\end{eqnarray}
Eq.~(\ref{eq:sepcl}) gives the frequency of the first type singular point for all the even order PCLs. Obviously when there exists odd(even) numbers of singular points for a certain band, the symmetry of the field inside layer A at the upper and lower band edge of the even order PCLs will be opposite(same).

For there exists an extra symmetry $\epsilon(\chi,-x)=\epsilon(d_b-\chi,x)$, we find:
\begin{equation}
E_n(K,x)=\pm E_{-n}(-K,-x)
\end{equation}
where n and $K$ denote the order of PCLs and Bloch vector respectively. So we have:
\begin{equation}
\label{eq:uf}
u_n(K,x)=\pm u_{-n}(-K,-x)
\end{equation}
where $u_n(K,x)$ is the periodic part of Bloch wave $E_n(K,x)=u_n(K,x)\exp(iKx)$. The Berry connection at the $K$ along the PCLs of order n is defined as: $A_K^n=\int{dx\epsilon_K^n(x)u_n^*(K,x)\partial_Ku_n(K,x)}$. Substituting Eq.~(\ref{eq:uf}) into $A_K^n$, we find
\begin{equation}
\label{eq:bc}
A_K^n=-A_{-K}^{-n}
\end{equation}
The Zak phase can be expressed as $\phi^n_{Zak}=\int_{-\pi/D}^{\pi/D} iA_K^n{dK}$\cite{PhysRevLett.62.2747}. Aside from the singularities, $\phi^{+n}$ and $\phi^{-n}$ cancel each out according to Eq.~(\ref{eq:bc}). So only the singularities contribute to the result of $\phi^{+n}+\phi^{-n}$. When there are even(odd) number singularities along PCLs of order n, the extended Zak phase defined in our paper $\phi^n_{Zak}=(\phi^{+n}+\phi^{-n})/2$ gives $0$($\pi$). Obviously, all the PCLs of even order have the same extended Zak phase, so do the PCLs of odd order. In conclusion, we state that the conclusions drawn in the system with SIS\cite{PhysRevX.4.021017} can be naturally extended to present 1D system with PWFDS without SIS(along the non-zero order PCLs) after redefinition of the Zak phase.

For the PCLs of odd order, the existence of the first type singularities is given by:
\begin{eqnarray}
&&\tan{k_cd_c}=\frac{F_3\sin{k_bd_b}}{F_4\sin^2{k_bd_b/2-F_2\cos^2{k_bd_b/2}}},\nonumber\\
&&\sin{k_cd_c}\neq0
\end{eqnarray}
Similar conclusion can be drawn for the odd order PCLs.

\section*{Appendix D: \label{D}The charge of the first type singularities in 2D parameter space}
To show the topology of the singularities in the 2D parameter space, we expand the $E^-=\exp(iKD)-T_{11}$ and $E^+=T_{12}\exp(ik_ad_a)$ around the singularity($K_0$, $\chi_0$) with respect to \{$\delta K$, $\delta \chi$\}.
As for a function $f=u(\omega, K, \chi)$
\begin{eqnarray}
&&\frac{\partial u}{\partial K}=\frac{\partial f}{\partial \omega}\frac{\partial \omega}{\partial K}+\frac{\partial f}{\partial K} \\ \nonumber\\
&&\frac{\partial u}{\partial \chi}=\frac{\partial f}{\partial \omega}\frac{\partial \omega}{\partial \chi}+\frac{\partial f}{\partial \chi}
\end{eqnarray}
The dispersion relation is determined by $F(\omega, K, \chi)=Re[T_{11}]-\cos{KD}=0$, so $\partial \omega/\partial K$, $\partial \omega/\partial \chi$ can be calculated by:
\begin{eqnarray}
&&\frac{\partial \omega}{\partial K}=\frac{-F'_K}{F'_\omega} \\ \nonumber\\
&&\frac{\partial \omega}{\partial \chi}=\frac{-F'_\chi}{F'_\omega}
\end{eqnarray}
We find:
\begin{eqnarray}
\exp(iKD)-T_{11} &&\propto \mathcal{O}(\delta K^2+\delta \chi^2) \\
T_{12}\exp(ik_ad_a) &&\propto \exp(ik_ad_a)(A\delta \chi+Bi\delta \omega)
\end{eqnarray}
where A, B are real number. The topological charge of the singularity is defined by the Berry phase around it $\oint(d\chi{A_\chi}+dK{A_K})/(2{\pi}i)$\cite{PhysRevLett.71.3697} which can be calculated by the winding number of the argument of $u(\omega, K, \chi)$ around it.
Since $u(\omega, K, \chi)$ can be express as a linear combination of $\exp(iKD)-T_{11}$ and $T_{12}\exp(ik_ad_a)$ and the singularity is located at the band branch with positive group velocity($\partial\omega/\partial{K}\geq0$) under our gauge, the winding number of the argument of $u(\omega, K, \chi)$ equals to the winding number of the argument of $T_{12}\exp(ik_ad_a)$ around the singularity. According to Eq.~(\ref{eq:rf}), the winding number of the argument of $T_{12}\exp(ik_ad_a)$ equals to the opposite of the reflection phase winding number in the reflection phase spectra. In conclusion, the first type singular points manifest as reflection phase vortex in the reflection phase spectra, whose topological charge equals to the opposite of the reflection phase winding number in the reflection phase spectra.

\section*{Appendix E: \label{E}The second type singularity and topological transition with nodal lines}
The condition for the second type singularity is given by:
\begin{equation}
\label{eq:sts}
\sin n_bd_b \omega'/c=0, \sin n_cd_c \omega'/c=0
\end{equation}
Substituting Eq.~(\ref{eq:sts}) into Eq.~(\ref{eq:t12}), it is easy to find that $T_{12}\exp(ik_ad_a)=0$ regardless of $\chi$. This second type singularities manifest as a singular line in the reflection phase spectra across which the reflection phase experiences $\pi$ phase jump and the reflection intensity reaches $1$ at the singular line. We assume the ratio of optical path of layers satisfies
\begin{equation}
n_a d_a:n_b d_b:n_c d_c=l:m:n
\end{equation}
where $l, m, n \in \mathbb{N^+}$. At the mid frequency $w_{l+m+n}=(l+m+n)\pi{c}/(n_ad_a+n_bd_b+n_cd_c)$ of the l+m+nth gap, we find that:
\begin{eqnarray}
|Re[T_{11}]|=1 \\
\partial ({Re[T_{11}]})/\partial {K}=0
\end{eqnarray}
So the two bands are degenerated at the frequency $w_{l+m+n}=(l+m+n)\pi{c}/(n_ad_a+n_bd_b+n_cd_c)$ regardless of $\chi$.
In conclusion when the singular line is tuned to the mid gap frequency $w_{l+m+n}=(l+m+n)\pi{c}/(n_ad_a+n_bd_b+n_cd_c)$ of the l+m+nth gap, the corresponding gap will turn into a nodal line at the mid gap frequency.
\bibliography{ref}

\providecommand{\noopsort}[1]{}\providecommand{\singleletter}[1]{#1}%
\begin{thebibliography}{26}%
\makeatletter
\providecommand \@ifxundefined [1]{%
 \@ifx{#1\undefined}
}%
\providecommand \@ifnum [1]{%
 \ifnum #1\expandafter \@firstoftwo
 \else \expandafter \@secondoftwo
 \fi
}%
\providecommand \@ifx [1]{%
 \ifx #1\expandafter \@firstoftwo
 \else \expandafter \@secondoftwo
 \fi
}%
\providecommand \natexlab [1]{#1}%
\providecommand \enquote  [1]{``#1''}%
\providecommand \bibnamefont  [1]{#1}%
\providecommand \bibfnamefont [1]{#1}%
\providecommand \citenamefont [1]{#1}%
\providecommand \href@noop [0]{\@secondoftwo}%
\providecommand \href [0]{\begingroup \@sanitize@url \@href}%
\providecommand \@href[1]{\@@startlink{#1}\@@href}%
\providecommand \@@href[1]{\endgroup#1\@@endlink}%
\providecommand \@sanitize@url [0]{\catcode `\\12\catcode `\$12\catcode
  `\&12\catcode `\#12\catcode `\^12\catcode `\_12\catcode `\%12\relax}%
\providecommand \@@startlink[1]{}%
\providecommand \@@endlink[0]{}%
\providecommand \url  [0]{\begingroup\@sanitize@url \@url }%
\providecommand \@url [1]{\endgroup\@href {#1}{\urlprefix }}%
\providecommand \urlprefix  [0]{URL }%
\providecommand \Eprint [0]{\href }%
\providecommand \doibase [0]{http://dx.doi.org/}%
\providecommand \selectlanguage [0]{\@gobble}%
\providecommand \bibinfo  [0]{\@secondoftwo}%
\providecommand \bibfield  [0]{\@secondoftwo}%
\providecommand \translation [1]{[#1]}%
\providecommand \BibitemOpen [0]{}%
\providecommand \bibitemStop [0]{}%
\providecommand \bibitemNoStop [0]{.\EOS\space}%
\providecommand \EOS [0]{\spacefactor3000\relax}%
\providecommand \BibitemShut  [1]{\csname bibitem#1\endcsname}%
\let\auto@bib@innerbib\@empty
\bibitem [{\citenamefont {Klitzing}\ \emph {et~al.}(1980)\citenamefont
  {Klitzing}, \citenamefont {Dorda},\ and\ \citenamefont
  {Pepper}}]{PhysRevLett.45.494}%
  \BibitemOpen
  \bibfield  {author} {\bibinfo {author} {\bibfnamefont {K.~v.}\ \bibnamefont
  {Klitzing}}, \bibinfo {author} {\bibfnamefont {G.}~\bibnamefont {Dorda}}, \
  and\ \bibinfo {author} {\bibfnamefont {M.}~\bibnamefont {Pepper}},\ }\href
  {\doibase 10.1103/PhysRevLett.45.494} {\bibfield  {journal} {\bibinfo
  {journal} {Phys. Rev. Lett.}\ }\textbf {\bibinfo {volume} {45}},\ \bibinfo
  {pages} {494} (\bibinfo {year} {1980})}\BibitemShut {NoStop}%
\bibitem [{\citenamefont {Qi}\ and\ \citenamefont
  {Zhang}(2011)}]{RevModPhys.83.1057}%
  \BibitemOpen
  \bibfield  {author} {\bibinfo {author} {\bibfnamefont {X.-L.}\ \bibnamefont
  {Qi}}\ and\ \bibinfo {author} {\bibfnamefont {S.-C.}\ \bibnamefont {Zhang}},\
  }\href {\doibase 10.1103/RevModPhys.83.1057} {\bibfield  {journal} {\bibinfo
  {journal} {Rev. Mod. Phys.}\ }\textbf {\bibinfo {volume} {83}},\ \bibinfo
  {pages} {1057} (\bibinfo {year} {2011})}\BibitemShut {NoStop}%
\bibitem [{\citenamefont {Hasan}\ and\ \citenamefont
  {Kane}(2010)}]{RevModPhys.82.3045}%
  \BibitemOpen
  \bibfield  {author} {\bibinfo {author} {\bibfnamefont {M.~Z.}\ \bibnamefont
  {Hasan}}\ and\ \bibinfo {author} {\bibfnamefont {C.~L.}\ \bibnamefont
  {Kane}},\ }\href {\doibase 10.1103/RevModPhys.82.3045} {\bibfield  {journal}
  {\bibinfo  {journal} {Rev. Mod. Phys.}\ }\textbf {\bibinfo {volume} {82}},\
  \bibinfo {pages} {3045} (\bibinfo {year} {2010})}\BibitemShut {NoStop}%
\bibitem [{\citenamefont {Lu}\ \emph {et~al.}(2014)\citenamefont {Lu},
  \citenamefont {Joannopoulos},\ and\ \citenamefont
  {Solja{\v{c}}i{\'c}}}]{luling}%
  \BibitemOpen
  \bibfield  {author} {\bibinfo {author} {\bibfnamefont {L.}~\bibnamefont
  {Lu}}, \bibinfo {author} {\bibfnamefont {J.~D.}\ \bibnamefont
  {Joannopoulos}}, \ and\ \bibinfo {author} {\bibfnamefont {M.}~\bibnamefont
  {Solja{\v{c}}i{\'c}}},\ }\href {\doibase 10.1038/nphoton.2014.248} {\bibfield
   {journal} {\bibinfo  {journal} {Nature Photonics}\ }\textbf {\bibinfo
  {volume} {8}},\ \bibinfo {pages} {821} (\bibinfo {year} {2014})}\BibitemShut
  {NoStop}%
\bibitem [{\citenamefont {Zak}(1989)}]{PhysRevLett.62.2747}%
  \BibitemOpen
  \bibfield  {author} {\bibinfo {author} {\bibfnamefont {J.}~\bibnamefont
  {Zak}},\ }\href {\doibase 10.1103/PhysRevLett.62.2747} {\bibfield  {journal}
  {\bibinfo  {journal} {Phys. Rev. Lett.}\ }\textbf {\bibinfo {volume} {62}},\
  \bibinfo {pages} {2747} (\bibinfo {year} {1989})}\BibitemShut {NoStop}%
\bibitem [{\citenamefont {Xiao}\ \emph {et~al.}(2014)\citenamefont {Xiao},
  \citenamefont {Zhang},\ and\ \citenamefont {Chan}}]{PhysRevX.4.021017}%
  \BibitemOpen
  \bibfield  {author} {\bibinfo {author} {\bibfnamefont {M.}~\bibnamefont
  {Xiao}}, \bibinfo {author} {\bibfnamefont {Z.~Q.}\ \bibnamefont {Zhang}}, \
  and\ \bibinfo {author} {\bibfnamefont {C.~T.}\ \bibnamefont {Chan}},\ }\href
  {\doibase 10.1103/PhysRevX.4.021017} {\bibfield  {journal} {\bibinfo
  {journal} {Phys. Rev. X}\ }\textbf {\bibinfo {volume} {4}},\ \bibinfo {pages}
  {021017} (\bibinfo {year} {2014})}\BibitemShut {NoStop}%
\bibitem [{\citenamefont {Zhu}\ \emph {et~al.}(2018{\natexlab{a}})\citenamefont
  {Zhu}, \citenamefont {Ding}, \citenamefont {Ren}, \citenamefont {Sun},
  \citenamefont {Li}, \citenamefont {Jiang},\ and\ \citenamefont {Chen}}]{abc}%
  \BibitemOpen
  \bibfield  {author} {\bibinfo {author} {\bibfnamefont {W.}~\bibnamefont
  {Zhu}}, \bibinfo {author} {\bibfnamefont {Y.-q.}\ \bibnamefont {Ding}},
  \bibinfo {author} {\bibfnamefont {J.}~\bibnamefont {Ren}}, \bibinfo {author}
  {\bibfnamefont {Y.}~\bibnamefont {Sun}}, \bibinfo {author} {\bibfnamefont
  {Y.}~\bibnamefont {Li}}, \bibinfo {author} {\bibfnamefont {H.}~\bibnamefont
  {Jiang}}, \ and\ \bibinfo {author} {\bibfnamefont {H.}~\bibnamefont {Chen}},\
  }\href {\doibase 10.1103/PhysRevB.97.195307} {\bibfield  {journal} {\bibinfo
  {journal} {Phys. Rev. B}\ }\textbf {\bibinfo {volume} {97}},\ \bibinfo
  {pages} {195307} (\bibinfo {year} {2018}{\natexlab{a}})}\BibitemShut
  {NoStop}%
\bibitem [{\citenamefont {Kalozoumis}\ \emph {et~al.}(2018)\citenamefont
  {Kalozoumis}, \citenamefont {Theocharis}, \citenamefont {Achilleos},
  \citenamefont {F\'elix}, \citenamefont {Richoux},\ and\ \citenamefont
  {Pagneux}}]{PhysRevA.98.023838}%
  \BibitemOpen
  \bibfield  {author} {\bibinfo {author} {\bibfnamefont {P.~A.}\ \bibnamefont
  {Kalozoumis}}, \bibinfo {author} {\bibfnamefont {G.}~\bibnamefont
  {Theocharis}}, \bibinfo {author} {\bibfnamefont {V.}~\bibnamefont
  {Achilleos}}, \bibinfo {author} {\bibfnamefont {S.}~\bibnamefont {F\'elix}},
  \bibinfo {author} {\bibfnamefont {O.}~\bibnamefont {Richoux}}, \ and\
  \bibinfo {author} {\bibfnamefont {V.}~\bibnamefont {Pagneux}},\ }\href
  {\doibase 10.1103/PhysRevA.98.023838} {\bibfield  {journal} {\bibinfo
  {journal} {Phys. Rev. A}\ }\textbf {\bibinfo {volume} {98}},\ \bibinfo
  {pages} {023838} (\bibinfo {year} {2018})}\BibitemShut {NoStop}%
\bibitem [{\citenamefont {Haldane}\ and\ \citenamefont
  {Raghu}(2008)}]{PhysRevLett.100.013904}%
  \BibitemOpen
  \bibfield  {author} {\bibinfo {author} {\bibfnamefont {F.~D.~M.}\
  \bibnamefont {Haldane}}\ and\ \bibinfo {author} {\bibfnamefont
  {S.}~\bibnamefont {Raghu}},\ }\href {\doibase 10.1103/PhysRevLett.100.013904}
  {\bibfield  {journal} {\bibinfo  {journal} {Phys. Rev. Lett.}\ }\textbf
  {\bibinfo {volume} {100}},\ \bibinfo {pages} {013904} (\bibinfo {year}
  {2008})}\BibitemShut {NoStop}%
\bibitem [{\citenamefont {Raghu}\ and\ \citenamefont
  {Haldane}(2008)}]{PhysRevA.78.033834}%
  \BibitemOpen
  \bibfield  {author} {\bibinfo {author} {\bibfnamefont {S.}~\bibnamefont
  {Raghu}}\ and\ \bibinfo {author} {\bibfnamefont {F.~D.~M.}\ \bibnamefont
  {Haldane}},\ }\href {\doibase 10.1103/PhysRevA.78.033834} {\bibfield
  {journal} {\bibinfo  {journal} {Phys. Rev. A}\ }\textbf {\bibinfo {volume}
  {78}},\ \bibinfo {pages} {033834} (\bibinfo {year} {2008})}\BibitemShut
  {NoStop}%
\bibitem [{\citenamefont {Gao}\ \emph {et~al.}(2017)\citenamefont {Gao},
  \citenamefont {Yang}, \citenamefont {Gao}, \citenamefont {Xue}, \citenamefont
  {Yang}, \citenamefont {Dong},\ and\ \citenamefont
  {Zhang}}]{PhysRevB.96.201402}%
  \BibitemOpen
  \bibfield  {author} {\bibinfo {author} {\bibfnamefont {Z.}~\bibnamefont
  {Gao}}, \bibinfo {author} {\bibfnamefont {Z.}~\bibnamefont {Yang}}, \bibinfo
  {author} {\bibfnamefont {F.}~\bibnamefont {Gao}}, \bibinfo {author}
  {\bibfnamefont {H.}~\bibnamefont {Xue}}, \bibinfo {author} {\bibfnamefont
  {Y.}~\bibnamefont {Yang}}, \bibinfo {author} {\bibfnamefont {J.}~\bibnamefont
  {Dong}}, \ and\ \bibinfo {author} {\bibfnamefont {B.}~\bibnamefont {Zhang}},\
  }\href {\doibase 10.1103/PhysRevB.96.201402} {\bibfield  {journal} {\bibinfo
  {journal} {Phys. Rev. B}\ }\textbf {\bibinfo {volume} {96}},\ \bibinfo
  {pages} {201402} (\bibinfo {year} {2017})}\BibitemShut {NoStop}%
\bibitem [{\citenamefont {Wu}\ and\ \citenamefont
  {Hu}(2015)}]{PhysRevLett.114.223901}%
  \BibitemOpen
  \bibfield  {author} {\bibinfo {author} {\bibfnamefont {L.-H.}\ \bibnamefont
  {Wu}}\ and\ \bibinfo {author} {\bibfnamefont {X.}~\bibnamefont {Hu}},\ }\href
  {\doibase 10.1103/PhysRevLett.114.223901} {\bibfield  {journal} {\bibinfo
  {journal} {Phys. Rev. Lett.}\ }\textbf {\bibinfo {volume} {114}},\ \bibinfo
  {pages} {223901} (\bibinfo {year} {2015})}\BibitemShut {NoStop}%
\bibitem [{\citenamefont {Hatsugai}(1993)}]{PhysRevLett.71.3697}%
  \BibitemOpen
  \bibfield  {author} {\bibinfo {author} {\bibfnamefont {Y.}~\bibnamefont
  {Hatsugai}},\ }\href {\doibase 10.1103/PhysRevLett.71.3697} {\bibfield
  {journal} {\bibinfo  {journal} {Phys. Rev. Lett.}\ }\textbf {\bibinfo
  {volume} {71}},\ \bibinfo {pages} {3697} (\bibinfo {year}
  {1993})}\BibitemShut {NoStop}%
\bibitem [{\citenamefont {Lu}\ \emph {et~al.}(2013)\citenamefont {Lu},
  \citenamefont {Fu}, \citenamefont {Joannopoulos},\ and\ \citenamefont
  {Solja{\v{c}}i{\'c}}}]{weyl}%
  \BibitemOpen
  \bibfield  {author} {\bibinfo {author} {\bibfnamefont {L.}~\bibnamefont
  {Lu}}, \bibinfo {author} {\bibfnamefont {L.}~\bibnamefont {Fu}}, \bibinfo
  {author} {\bibfnamefont {J.~D.}\ \bibnamefont {Joannopoulos}}, \ and\
  \bibinfo {author} {\bibfnamefont {M.}~\bibnamefont {Solja{\v{c}}i{\'c}}},\
  }\href {\doibase 10.1038/nphoton.2013.42
  https://www.nature.com/articles/nphoton.2013.42#supplementary-information}
  {\bibfield  {journal} {\bibinfo  {journal} {Nature Photonics}\ }\textbf
  {\bibinfo {volume} {7}},\ \bibinfo {pages} {294} (\bibinfo {year}
  {2013})}\BibitemShut {NoStop}%
\bibitem [{\citenamefont {Xiao}\ \emph {et~al.}(2016)\citenamefont {Xiao},
  \citenamefont {Lin},\ and\ \citenamefont {Fan}}]{PhysRevLett.117.057401}%
  \BibitemOpen
  \bibfield  {author} {\bibinfo {author} {\bibfnamefont {M.}~\bibnamefont
  {Xiao}}, \bibinfo {author} {\bibfnamefont {Q.}~\bibnamefont {Lin}}, \ and\
  \bibinfo {author} {\bibfnamefont {S.}~\bibnamefont {Fan}},\ }\href {\doibase
  10.1103/PhysRevLett.117.057401} {\bibfield  {journal} {\bibinfo  {journal}
  {Phys. Rev. Lett.}\ }\textbf {\bibinfo {volume} {117}},\ \bibinfo {pages}
  {057401} (\bibinfo {year} {2016})}\BibitemShut {NoStop}%
\bibitem [{\citenamefont {Chang}\ \emph {et~al.}(2017)\citenamefont {Chang},
  \citenamefont {Xiao}, \citenamefont {Chen},\ and\ \citenamefont
  {Chan}}]{PhysRevB.95.125136}%
  \BibitemOpen
  \bibfield  {author} {\bibinfo {author} {\bibfnamefont {M.-L.}\ \bibnamefont
  {Chang}}, \bibinfo {author} {\bibfnamefont {M.}~\bibnamefont {Xiao}},
  \bibinfo {author} {\bibfnamefont {W.-J.}\ \bibnamefont {Chen}}, \ and\
  \bibinfo {author} {\bibfnamefont {C.~T.}\ \bibnamefont {Chan}},\ }\href
  {\doibase 10.1103/PhysRevB.95.125136} {\bibfield  {journal} {\bibinfo
  {journal} {Phys. Rev. B}\ }\textbf {\bibinfo {volume} {95}},\ \bibinfo
  {pages} {125136} (\bibinfo {year} {2017})}\BibitemShut {NoStop}%
\bibitem [{\citenamefont {Wang}\ \emph {et~al.}(2017)\citenamefont {Wang},
  \citenamefont {Xiao}, \citenamefont {Liu}, \citenamefont {Zhu},\ and\
  \citenamefont {Chan}}]{PhysRevX.7.031032}%
  \BibitemOpen
  \bibfield  {author} {\bibinfo {author} {\bibfnamefont {Q.}~\bibnamefont
  {Wang}}, \bibinfo {author} {\bibfnamefont {M.}~\bibnamefont {Xiao}}, \bibinfo
  {author} {\bibfnamefont {H.}~\bibnamefont {Liu}}, \bibinfo {author}
  {\bibfnamefont {S.}~\bibnamefont {Zhu}}, \ and\ \bibinfo {author}
  {\bibfnamefont {C.~T.}\ \bibnamefont {Chan}},\ }\href {\doibase
  10.1103/PhysRevX.7.031032} {\bibfield  {journal} {\bibinfo  {journal} {Phys.
  Rev. X}\ }\textbf {\bibinfo {volume} {7}},\ \bibinfo {pages} {031032}
  (\bibinfo {year} {2017})}\BibitemShut {NoStop}%
\bibitem [{\citenamefont {Roushan}\ \emph {et~al.}(2014)\citenamefont
  {Roushan}, \citenamefont {Neill}, \citenamefont {Chen}, \citenamefont
  {Kolodrubetz}, \citenamefont {Quintana}, \citenamefont {Leung}, \citenamefont
  {Fang}, \citenamefont {Barends}, \citenamefont {Campbell}, \citenamefont
  {Chen}, \citenamefont {Chiaro}, \citenamefont {Dunsworth}, \citenamefont
  {Jeffrey}, \citenamefont {Kelly}, \citenamefont {Megrant}, \citenamefont
  {Mutus}, \citenamefont {O¡¯Malley}, \citenamefont {Sank}, \citenamefont
  {Vainsencher}, \citenamefont {Wenner}, \citenamefont {White}, \citenamefont
  {Polkovnikov}, \citenamefont {Cleland},\ and\ \citenamefont
  {Martinis}}]{Observation}%
  \BibitemOpen
  \bibfield  {author} {\bibinfo {author} {\bibfnamefont {P.}~\bibnamefont
  {Roushan}}, \bibinfo {author} {\bibfnamefont {C.}~\bibnamefont {Neill}},
  \bibinfo {author} {\bibfnamefont {Y.}~\bibnamefont {Chen}}, \bibinfo {author}
  {\bibfnamefont {M.}~\bibnamefont {Kolodrubetz}}, \bibinfo {author}
  {\bibfnamefont {C.}~\bibnamefont {Quintana}}, \bibinfo {author}
  {\bibfnamefont {N.}~\bibnamefont {Leung}}, \bibinfo {author} {\bibfnamefont
  {M.}~\bibnamefont {Fang}}, \bibinfo {author} {\bibfnamefont {R.}~\bibnamefont
  {Barends}}, \bibinfo {author} {\bibfnamefont {B.}~\bibnamefont {Campbell}},
  \bibinfo {author} {\bibfnamefont {Z.}~\bibnamefont {Chen}}, \bibinfo {author}
  {\bibfnamefont {B.}~\bibnamefont {Chiaro}}, \bibinfo {author} {\bibfnamefont
  {A.}~\bibnamefont {Dunsworth}}, \bibinfo {author} {\bibfnamefont
  {E.}~\bibnamefont {Jeffrey}}, \bibinfo {author} {\bibfnamefont
  {J.}~\bibnamefont {Kelly}}, \bibinfo {author} {\bibfnamefont
  {A.}~\bibnamefont {Megrant}}, \bibinfo {author} {\bibfnamefont
  {J.}~\bibnamefont {Mutus}}, \bibinfo {author} {\bibfnamefont {P.~J.~J.}\
  \bibnamefont {O¡¯Malley}}, \bibinfo {author} {\bibfnamefont {D.}~\bibnamefont
  {Sank}}, \bibinfo {author} {\bibfnamefont {A.}~\bibnamefont {Vainsencher}},
  \bibinfo {author} {\bibfnamefont {J.}~\bibnamefont {Wenner}}, \bibinfo
  {author} {\bibfnamefont {T.}~\bibnamefont {White}}, \bibinfo {author}
  {\bibfnamefont {A.}~\bibnamefont {Polkovnikov}}, \bibinfo {author}
  {\bibfnamefont {A.~N.}\ \bibnamefont {Cleland}}, \ and\ \bibinfo {author}
  {\bibfnamefont {J.~M.}\ \bibnamefont {Martinis}},\ }\href {\doibase
  10.1038/nature13891} {\bibfield  {journal} {\bibinfo  {journal} {Nature}\
  }\textbf {\bibinfo {volume} {515}},\ \bibinfo {pages} {241} (\bibinfo {year}
  {2014})}\BibitemShut {NoStop}%
\bibitem [{\citenamefont {Schroer}\ \emph {et~al.}(2014)\citenamefont
  {Schroer}, \citenamefont {Kolodrubetz}, \citenamefont {Kindel}, \citenamefont
  {Sandberg}, \citenamefont {Gao}, \citenamefont {Vissers}, \citenamefont
  {Pappas}, \citenamefont {Polkovnikov},\ and\ \citenamefont
  {Lehnert}}]{PhysRevLett.113.050402}%
  \BibitemOpen
  \bibfield  {author} {\bibinfo {author} {\bibfnamefont {M.~D.}\ \bibnamefont
  {Schroer}}, \bibinfo {author} {\bibfnamefont {M.~H.}\ \bibnamefont
  {Kolodrubetz}}, \bibinfo {author} {\bibfnamefont {W.~F.}\ \bibnamefont
  {Kindel}}, \bibinfo {author} {\bibfnamefont {M.}~\bibnamefont {Sandberg}},
  \bibinfo {author} {\bibfnamefont {J.}~\bibnamefont {Gao}}, \bibinfo {author}
  {\bibfnamefont {M.~R.}\ \bibnamefont {Vissers}}, \bibinfo {author}
  {\bibfnamefont {D.~P.}\ \bibnamefont {Pappas}}, \bibinfo {author}
  {\bibfnamefont {A.}~\bibnamefont {Polkovnikov}}, \ and\ \bibinfo {author}
  {\bibfnamefont {K.~W.}\ \bibnamefont {Lehnert}},\ }\href {\doibase
  10.1103/PhysRevLett.113.050402} {\bibfield  {journal} {\bibinfo  {journal}
  {Phys. Rev. Lett.}\ }\textbf {\bibinfo {volume} {113}},\ \bibinfo {pages}
  {050402} (\bibinfo {year} {2014})}\BibitemShut {NoStop}%
\bibitem [{\citenamefont {Zhu}\ \emph {et~al.}(2018{\natexlab{b}})\citenamefont
  {Zhu}, \citenamefont {Fang}, \citenamefont {Li}, \citenamefont {Sun},
  \citenamefont {Li}, \citenamefont {Jing},\ and\ \citenamefont
  {Chen}}]{PhysRevLett.121.124501}%
  \BibitemOpen
  \bibfield  {author} {\bibinfo {author} {\bibfnamefont {W.}~\bibnamefont
  {Zhu}}, \bibinfo {author} {\bibfnamefont {X.}~\bibnamefont {Fang}}, \bibinfo
  {author} {\bibfnamefont {D.}~\bibnamefont {Li}}, \bibinfo {author}
  {\bibfnamefont {Y.}~\bibnamefont {Sun}}, \bibinfo {author} {\bibfnamefont
  {Y.}~\bibnamefont {Li}}, \bibinfo {author} {\bibfnamefont {Y.}~\bibnamefont
  {Jing}}, \ and\ \bibinfo {author} {\bibfnamefont {H.}~\bibnamefont {Chen}},\
  }\href {\doibase 10.1103/PhysRevLett.121.124501} {\bibfield  {journal}
  {\bibinfo  {journal} {Phys. Rev. Lett.}\ }\textbf {\bibinfo {volume} {121}},\
  \bibinfo {pages} {124501} (\bibinfo {year} {2018}{\natexlab{b}})}\BibitemShut
  {NoStop}%
\bibitem [{\citenamefont {Lang}\ \emph {et~al.}(2012)\citenamefont {Lang},
  \citenamefont {Cai},\ and\ \citenamefont {Chen}}]{PhysRevLett.108.220401}%
  \BibitemOpen
  \bibfield  {author} {\bibinfo {author} {\bibfnamefont {L.-J.}\ \bibnamefont
  {Lang}}, \bibinfo {author} {\bibfnamefont {X.}~\bibnamefont {Cai}}, \ and\
  \bibinfo {author} {\bibfnamefont {S.}~\bibnamefont {Chen}},\ }\href {\doibase
  10.1103/PhysRevLett.108.220401} {\bibfield  {journal} {\bibinfo  {journal}
  {Phys. Rev. Lett.}\ }\textbf {\bibinfo {volume} {108}},\ \bibinfo {pages}
  {220401} (\bibinfo {year} {2012})}\BibitemShut {NoStop}%
\bibitem [{\citenamefont {Poshakinskiy}\ \emph {et~al.}(2014)\citenamefont
  {Poshakinskiy}, \citenamefont {Poddubny}, \citenamefont {Pilozzi},\ and\
  \citenamefont {Ivchenko}}]{PhysRevLett.112.107403}%
  \BibitemOpen
  \bibfield  {author} {\bibinfo {author} {\bibfnamefont {A.~V.}\ \bibnamefont
  {Poshakinskiy}}, \bibinfo {author} {\bibfnamefont {A.~N.}\ \bibnamefont
  {Poddubny}}, \bibinfo {author} {\bibfnamefont {L.}~\bibnamefont {Pilozzi}}, \
  and\ \bibinfo {author} {\bibfnamefont {E.~L.}\ \bibnamefont {Ivchenko}},\
  }\href {\doibase 10.1103/PhysRevLett.112.107403} {\bibfield  {journal}
  {\bibinfo  {journal} {Phys. Rev. Lett.}\ }\textbf {\bibinfo {volume} {112}},\
  \bibinfo {pages} {107403} (\bibinfo {year} {2014})}\BibitemShut {NoStop}%
\bibitem [{\citenamefont {Poshakinskiy}\ \emph {et~al.}(2015)\citenamefont
  {Poshakinskiy}, \citenamefont {Poddubny},\ and\ \citenamefont
  {Hafezi}}]{PhysRevA.91.043830}%
  \BibitemOpen
  \bibfield  {author} {\bibinfo {author} {\bibfnamefont {A.~V.}\ \bibnamefont
  {Poshakinskiy}}, \bibinfo {author} {\bibfnamefont {A.~N.}\ \bibnamefont
  {Poddubny}}, \ and\ \bibinfo {author} {\bibfnamefont {M.}~\bibnamefont
  {Hafezi}},\ }\href {\doibase 10.1103/PhysRevA.91.043830} {\bibfield
  {journal} {\bibinfo  {journal} {Phys. Rev. A}\ }\textbf {\bibinfo {volume}
  {91}},\ \bibinfo {pages} {043830} (\bibinfo {year} {2015})}\BibitemShut
  {NoStop}%
\bibitem [{\citenamefont {Pilozzi}\ and\ \citenamefont
  {Conti}(2016)}]{PhysRevB.93.195317}%
  \BibitemOpen
  \bibfield  {author} {\bibinfo {author} {\bibfnamefont {L.}~\bibnamefont
  {Pilozzi}}\ and\ \bibinfo {author} {\bibfnamefont {C.}~\bibnamefont
  {Conti}},\ }\href {\doibase 10.1103/PhysRevB.93.195317} {\bibfield  {journal}
  {\bibinfo  {journal} {Phys. Rev. B}\ }\textbf {\bibinfo {volume} {93}},\
  \bibinfo {pages} {195317} (\bibinfo {year} {2016})}\BibitemShut {NoStop}%
\bibitem [{\citenamefont {{Li}}\ and\ \citenamefont
  {{Jiang}}(2018)}]{2018arXiv181012550L}%
  \BibitemOpen
  \bibfield  {author} {\bibinfo {author} {\bibfnamefont {Q.}~\bibnamefont
  {{Li}}}\ and\ \bibinfo {author} {\bibfnamefont {X.}~\bibnamefont {{Jiang}}},\
  }\href@noop {} {\bibfield  {journal} {\bibinfo  {journal} {ArXiv e-prints}\ }
  (\bibinfo {year} {2018})},\ \Eprint {http://arxiv.org/abs/1810.12550}
  {arXiv:1810.12550 [physics.optics]} \BibitemShut {NoStop}%
\bibitem [{\citenamefont {Ezawa}(2013)}]{PhysRevB.87.155415}%
  \BibitemOpen
  \bibfield  {author} {\bibinfo {author} {\bibfnamefont {M.}~\bibnamefont
  {Ezawa}},\ }\href {\doibase 10.1103/PhysRevB.87.155415} {\bibfield  {journal}
  {\bibinfo  {journal} {Phys. Rev. B}\ }\textbf {\bibinfo {volume} {87}},\
  \bibinfo {pages} {155415} (\bibinfo {year} {2013})}\BibitemShut {NoStop}%
\end{thebibliography}%
\end{document}